\documentclass[journal]{IEEEtran}

\usepackage{graphicx}
\usepackage{epsfig}
\usepackage{amstext}
\usepackage{amssymb}
\usepackage{amsmath}
\usepackage[font=footnotesize]{caption}
\usepackage{subcaption}
\usepackage{cite}
\usepackage{float}
\usepackage{array}
\usepackage{url}

\usepackage{soul}
\usepackage{lipsum}
\usepackage{latexsym}
\usepackage{dcolumn}
\usepackage{widetext}
\usepackage{cuted}
\usepackage[ruled,vlined,linesnumbered]{algorithm2e}
\usepackage{colortbl}

\newcounter{mytempeqncnt}
\newif\ifconf
\conffalse

\IEEEoverridecommandlockouts
\DeclareMathOperator*{\argmax}{arg\,max}
\DeclareMathOperator*{\argmin}{arg\,min}

\begin{document}

\title{User Selection and Power Allocation in \\ Full Duplex Multi-Cell Networks}

\ifconf
    \author{\IEEEauthorblockN{Sanjay Goyal,
    Pei Liu,
    Shivendra S. Panwar} \\
    \IEEEauthorblockA{Department of Electrical and Computer Engineering\\
     NYU Tandon School of Engineering,
    Brooklyn, New York 11201\\ Email:\{sanjay.goyal,peiliu,panwar\}@nyu.edu}
}

\else
    \author{
        Sanjay Goyal,~\IEEEmembership{Student Member,~IEEE},
        Pei Liu,~\IEEEmembership{Member,~IEEE},
        and Shivendra S. Panwar,~\IEEEmembership{Fellow,~IEEE}
        \thanks{Copyright (c) 2015 IEEE. Personal use of this material is permitted. However, permission to use this material for any other purposes must be obtained from the IEEE by sending a request to pubs-permissions@ieee.org.}
        \thanks{This material is based upon work supported by the National Science Foundation under Grant No. 1527750, as well as generous support from the NYSTAR Center for Advanced Technology in Telecommunications (CATT), and InterDigital Communications.}
        \thanks{
            S. Goyal (email: sanjay.goyal@nyu.edu),
            P. Liu (email: peiliu@nyu.edu),
            S. S. Panwar (email: panwar@nyu.edu),
            are with the ECE Department, NYU Tandon School of Engineering,
            New York University, Brooklyn, NY.}
    }
\fi

\maketitle

\begin{abstract}
Full duplex (FD) communications has the potential to double the capacity of a half duplex (HD) system at the link level. However, in a cellular network, FD operation is not a straightforward extension of half duplex operations. The increased interference due to a large number of simultaneous transmissions in FD operation and realtime traffic conditions limits the capacity improvement. Realizing the potential of FD requires careful coordination of resource allocation among the cells as well as within the cell. In this paper, we propose a distributed resource allocation, i.e., joint user selection and power allocation for a FD multi-cell system, assuming FD base stations (BSs) and HD user equipment (UEs). Due to the complexity of finding the globally optimum solution, a sub-optimal solution for UE selection, and a novel geometric programming based solution for power allocation, are proposed. The proposed distributed approach converges quickly and performs almost as well as a centralized solution, but with much lower signaling overhead. It provides a hybrid scheduling policy which allows FD operations whenever it is advantageous, but otherwise defaults to HD operation. We focus on small cell systems because they are more suitable for FD operation, given practical self-interference cancellation limits. With practical self-interference cancellation, it is shown that the proposed hybrid FD system achieves nearly two times throughput improvement for an indoor multi-cell scenario, and about 65$\%$ improvement for an outdoor multi-cell scenario compared to the HD system.

\end{abstract}
\begin{IEEEkeywords}
Full duplex radio, LTE, small cell, scheduling, power allocation.
\end{IEEEkeywords}

\section{Introduction}\label{sec:intro}

\IEEEPARstart{F}{ull} duplex (FD) operation in a single wireless channel has the potential to double the spectral efficiency of a wireless point-to-point link by transmitting in both directions at the same time. Motivated by the rapid growth in wireless data traffic, along with concerns about a spectrum shortage \cite{Cisco,Ericsson,Horizon}, cellular network operators and system vendors have become more interested in FD operations. 

In legacy systems, the large difference between transmitted (Tx) and received (Rx) signal powers due to path loss and fading, together with imperfect Tx/Rx isolation, has driven the vast majority of systems to use either frequency division duplexing (FDD) or time division duplexing (TDD). FDD separates the Tx and Rx signals with filters while TDD achieves this with Tx/Rx switching. Recent advances in antenna designs and active cancellation technologies~\cite{Khandani10, Katti10, Knox12, Katti13, Duarte13,survey_JSAC,survey_kim} provide a significant step towards building a practical FD transceiver and meeting the projected 2X gain in capacity~\cite{NGMN_5G,kumu_5G} without requiring new spectrum or setting up new cells. A combination of antenna, analog and digital cancellation circuits can remove most of the crosstalk, or self-interference, between the Tx/Rx signal path, and allows demodulation of the received signal while transmitting to someone else. This was demonstrated using multiple antennas positioned for optimum cancellation~\cite{Khandani10,Katti10}, and later for single antenna systems~\cite{Knox12,Katti13}, where as much as 110 dB cancellation is reported over an 80 MHz bandwidth. Cancellation ranging from 70 to 100 dB with a median of 85 dB using multiple antennas has been reported~\cite{Duarte13}. An antenna feed network, for which a prototype provided 40 to 45 dB Tx/Rx isolation before analog and digital cancellation, was described in~\cite{Knox12}. %More references related to other FD circuit designs can be found in survey papers \cite{survey_JSAC,survey_kim}.

However, at the network level, FD operations in a cellular network is not just a straightforward extension of half duplex (HD) operations implemented by replacing the HD radios with a FD radio. As suggested in our preliminary research for LTE systems~\cite{SanjayCISS13,SanjayICC14} and by others~\cite{ChoiSTR12,XShen13,HyunICTC,multi_cell_GPonly}, intra/inter-cell interference caused by using the same frequency in both uplink and downlink directions is significant, and is a major limiting factor to the system throughput. This is becoming a key problem to resolve as new cellular networks become more heterogeneous, and network entities with different capabilities are loosely connected with each other. Additionally, realistic traffic complicates scheduling decisions since the scheduled user equipment (UE) might only have active traffic in one direction at a given instant. In such a scenario, it is advantageous to schedule a second UE in the opposite direction.

\begin{figure*}
\centering
\includegraphics[width = 5.5in] {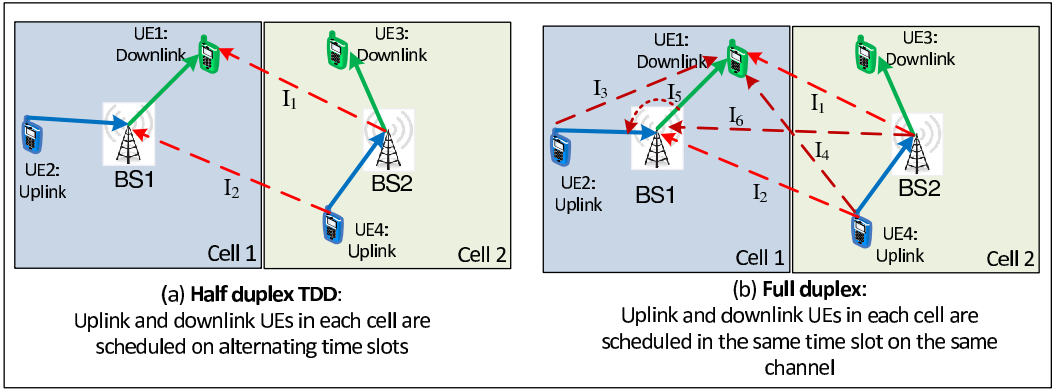}
\caption{Half duplex and full duplex multi-cell interference scenarios.}
\label{fig:fig1}
\end{figure*}

In this paper, we assume the BSs are equipped with FD radios, where the additional cost and power is most likely to be acceptable; while the UE is limited to HD operation. During FD operation in a cell, the BS schedules an uplink UE and a downlink UE in the same time slot on the same channel. The impact on over-the-air interference due to FD operation is illustrated in Figure~\ref{fig:fig1}. Consider the two-cell network in Figure~\ref{fig:fig1}, in which UE1 and UE3 are downlink UEs in cell 1 and cell 2, respectively, and UE2 and UE4 are uplink UEs in cell 1 and cell 2, respectively. First, to illustrate the HD scenario, we assume synchronized cells, which means that in a given time interval all cells schedule transmissions in the same direction. In this case, orthogonal channel access in time prevents interference between UEs and between base stations (BSs), but each UE accesses the channel only half the time. From Figure~\ref{fig:fig1}(a) one can see that in HD operation, UE1 receives interference ($I_1$) from BS2, which is transmitting to UE3 at the same time. Similarly, BS1 receives interference ($I_2$) from the uplink signal of UE4. During FD operation, as shown in Figure~\ref{fig:fig1}(b), the downlink UE, UE1, not only gets interference ($I_1$) from BS2, but also gets interference ($I_3$ and $I_4$) from the uplink signals of UE2 and UE4. Similarly, the uplink from UE2 to BS1 not only gets interference ($I_2$) from UE4, but also gets interference ($I_6$) from the downlink signal of BS2, as well as Tx-to-Rx self-interference ($I_5$). The existence of additional interference sources raises the question whether there is any net capacity gain from FD operation. The actual gain from FD operation will strongly depend on link geometries, the density of UEs, and propagation effects in mobile channels. Therefore, FD operation will provide a net throughput gain only if the throughput across two time slots, subject to the additional interference, is larger than the throughput in one time slot without such interference.

In this paper, we focus on the design of a distributed, interference-aware scheduler and power control algorithm that maximizes the FD gain across multiple cells, while maintaining a level of fairness between all UEs.  In such a system, FD gain can be achieved by simultaneous transmissions in uplink and downlink directions, where the the extra FD interference would be treated as noise. The scheduler is a hybrid scheduler in the sense that it will exploit FD transmissions at the BS only when it is advantageous to do so. Otherwise, when the interference is too strong, or traffic demands dictate it, it might conduct HD operations in some cells.

In the proposed distributed approach, neighboring cells coordinate with each other to simultaneously select the UEs and transmit power levels to maximize the system gain. This joint UE selection and power allocation problem is in general a non-convex, nonlinear, and mixed discrete optimization problem. There exists no method to find a globally optimum solution for such a problem, even for the traditional HD system scenario. We provide a sub-optimal method by separating the UE selection and power allocation procedures, using Geometric Programming (GP) for power allocation. The proposed distributed approach converges quickly and performs almost as well as a centralized solution which has access to global information, i.e., channel state information, power, etc., with much lower signaling overhead. The proposed FD system improves the capacity of a dense indoor multi-cell system by nearly two times and an outdoor sparse multi-cell system by about 65\%. The new signaling requirements and its overhead in the case of the FD scheduling process are also discussed. 

\subsection{Related Work} 
Extensive advances have been made in designing and implementing wireless transceivers with FD capability \cite{survey_JSAC,survey_kim}. MAC designs for FD IEEE 802.11 systems have been presented which shows throughput gains from 1.2x to 2.0x with FD operations (please refer to~\cite{survey_WLAN} and references therein). However, to the best of our knowledge, little has been done to understand the impact of such terminals on a cellular network in terms of system capacity and energy efficiency. 

Reviewing the literature shows that there has been significant work done on interference coordination in conventional HD systems. Various solutions~\cite{survey_IC_downlink} have been proposed from static frequency allocation to dynamic distributed resource allocation to avoid or coordinate the interference among neighboring interfering cells. However, with the new FD interference as described in Figure~\ref{fig:fig1}, uplink and downlink channel resources have to be allocated jointly to support a higher number of simultaneous links with different characteristics. Thus, the existing interference coordination methods for the HD case cannot be applied directly to the FD case. 

FD operation in a single cell has been evaluated\cite{Barghi12,SanjayICC14,DiINFOCOM,Shaocommletter,GMSingleCell,ouyang2015leveraging,Nguyenduplo}. Barghi \emph{et al.}~\cite{Barghi12} compared the tradeoff between using multiple antennas for spatial multiplexing gain and FD gain by nulling self-interference.  A distributed power control method using just one hop information to manage UE-to-UE interference in a single FD cell with massive MIMO was proposed in~\cite{ouyang2015leveraging}. FD operation in a cellular system has also been investigated in the DUPLO project~\cite{duplo_site}, where a joint uplink-downlink beamforming technique was designed for the single small cell environment~\cite{Nguyenduplo}. Our previous work~\cite{SanjayICC14} introduced a single cell hybrid scheduler without transmission power optimization. Other techniques for resource allocation in a FD single cell case using matching theory, a cell partitioning method, and game theory can be found in~\cite{DiINFOCOM},~\cite{Shaocommletter}, and~\cite{GMSingleCell}, respectively. However all these proposed methods for single FD cell cannot be directly applied to resource allocation in a multi-cell scenario. 

In the case of multi-cell FD operations, centralized UE selection procedures with fixed power allocation have been proposed \cite{XShen13, HyunICTC, SanjayCISS13}. Moreover, inter-BS interference is assumed to be perfectly cancelled and the interference from the neighboring cell UE is ignored in~\cite{XShen13,HyunICTC}, which makes the resource allocation problem simpler even for the multi-cell case. Under the same assumption, an analytical expression for the achievable rates assuming Cloud Radio Access Network (C-RAN) operation for both HD and FD are derived by Simeone \emph{et al.}\cite{Simeone2014full}. However, the assumption of ignoring interference from UEs of neighboring cells may not be appropriate in some scenarios. A cell-edge uplink UE of a neighboring cell may generate severe interference for the downlink transmission. Choi \emph{et al.}~\cite{ChoiSTR12} proposed a method to mitigate the inter-BS interference using null forming in the elevation angle at BS antennas and a simple UE selection procedure by assuming fixed transmission powers in both directions. Using successive convex approximation and GP, Nguyen \emph{et al.}~\cite{multi_cell_GPonly} provides a centralized power allocation method for the given UEs with FD capability. Yun \emph{et al.}~\cite{multi_cell_yun} provided a intra-cell joint resource allocation including channel allocation, UE selection, and power allocation. Further, they considered a multi femto-cell network with an underlying macro cell, for which they provided a coordination algorithm such that the transmit powers of femtocells and their connected UEs are adjusted so that data transmissions of the underlying macrocell is protected. However, they did not consider coordination to mitigate the interference among the co-channel femto cells. A high level presentation, without any technical details, of the centralized solution we use as an upper bound has been given in~\cite{Goyal_CommMag}, which was used to evaluate the performance of FD systems in an indoor multi-cell system in terms of energy efficiency. The details of this centralized method will be provided in Section~\ref{sec:CFDMR}.

Stochastic geometry based analytical models have also been presented \cite{SanjayCISS13}\cite{alves2014average, Quekhybrid, Goyal_ICC16} for the FD multi-cell system. The impact of residual self-interference, density of FD BSs, transmit power, etc., on the performance of such FD system in terms of average spectral efficiency and coverage has been evaluated. These stochastic geometry based analyses do not consider multi-UE diversity gain, which comes through scheduling of the appropriate UEs with power adjustments to mitigate interference. This is especially crucial in FD systems where, as we have just noted, the interference scenario is worse than traditional HD systems.

In this paper we provide a distributed method of interference coordination between cells with the appropriate UE selection and power allocation for a FD enabled cellular system. The key contributions of this paper are:
\begin{itemize}
\item A joint uplink and downlink scheduler is introduced, which maximizes network utility for a FD enabled multi-cell network.
\item The scheduler jointly optimizes UE selection and power allocation among multiple cells in a distributed manner.  
\item New signaling required to avoid UE-to-UE interference is discussed. The signaling overhead is also illustrated.
\item The paper investigates the performance of FD operations for several typical deployments used by cellular operators today, including both indoor and outdoor scenarios.
\end{itemize}

The remaining part of the paper is organized as follows: Section~\ref{sec:SMPF} describes the system model and problem formulation. The discussion on new requirements for channel estimation is discussed in Section~\ref{sec:CEP}. The distributed joint UE selection and power allocation method is given in Section~\ref{sec:DFDMR}. Section~\ref{sec:CFDMR} gives the details of a centralized method to solve the same problem. Section~\ref{sec:PE} contains simulation details and performance results for the proposed FD scheduling algorithms. Conclusions are discussed in Section~\ref{sec:conc}.

\section{System Description and Problem Formulation}\label{sec:SMPF}

\subsection{System Model}\label{subsec:SMPFS1}
We examine FD common carrier operation applied to a resource managed LTE TDD small-cell system~\cite{DahlmanLTE, 3GPP:2}. Residual self-interference, in general, lowers the uplink coverage and precludes the use of FD technology in a large cell. For example, consider a cell with a 1 kilometer radius. According to the channel model given in~\cite{3GPP:1}, the path loss at the cell edge is around 130 dB. It means the uplink signal arriving at the BS is 130 dB lower than the downlink signal transmitted, assuming equal per channel transmission power in the uplink and downlink directions. The received signal to interference ratio (SIR) will then be at most -20 dB with the best self-interference cancellation circuit known to date, which is capable of achieving 110 dB of cancellation~\cite{Katti13}. At such an SIR, the spectrum efficiency would be very low. Thus we believe FD transmission is more suitable for UEs close to base stations, which motivates us to consider small-cell systems as more suitable candidates to deploy an FD BS. 

We consider a network with $M$ cells, where $\Pi$ will be used to denote the set of indices of all BSs/cells. Each UE is connected to the nearest BS, and the number of UEs is much larger than $M$. We denote by $\mathcal{K}_m$ the set of UE indices associated with cell $m$, and define $N_m = |\mathcal{K}_m|$. Each of the BSs and UE devices are equipped with a single antenna. 

Assume that at timeslot $t$, $\psi_b^d(t) \in \mathcal{K}_b$ and $\psi_b^u(t) \in \mathcal{K}_b$ denote the UEs scheduled in cell $b$ in downlink and uplink directions, respectively. In case of HD UEs, $\psi_b^d(t) \neq \psi_b^u(t)$. The baseband signal received by UEs $\psi_b^d(t)$ and $\psi_b^u(t)$ are given by, respectively, 
\begin{equation}\label{SM1}
\begin{split}
y_{\psi_b^d(t)} (t) = \underbrace{h_{b,\psi_b^d(t)}~x_b(t)}_\text{data} + \underbrace{\sum_{i \in \Pi\setminus b} {h_{i, \psi_b^d(t)}~x_i(t)}}_\text{BS-to-UE interference} \\
+ \underbrace{\sum_{i \in \Pi} {h_{\psi_i^u(t), \psi_b^d(t)}~x_{\psi_i^u(t)}(t)}}_\text{UE-to-UE interference} + \underbrace{n_{\psi_b^d(t)}}_\text{noise},
\end{split}
\end{equation}

\begin{equation}\label{SM2}
\begin{split}
y_{\psi_b^u(t)} (t) = \underbrace{h_{\psi_b^u(t),b}~x_{\psi_b^u(t)}(t)}_\text{data} + \underbrace{\sum_{i \in \Pi\setminus b} {h_{\psi_i^u(t),b}~x_{\psi_i^u(t)}(t)}}_\text{UE-to-BS interference} \\
+ \underbrace{\sum_{i \in \Pi\setminus b} {h_{i,b}~x_i(t)}}_\text{BS-to-BS interference} + \underbrace{h'_{b,b}~x_b}_\text{self-interference} + \underbrace{n_{b}}_\text{noise}.
\end{split}
\end{equation}

In the above equations, $h_{\{\}}$ is used to denote the complex channel response between different nodes. For example, $h_{b,\psi_b^d(t)}$ and $h_{\psi_i^u(t), \psi_b^d(t)}$ denote the channel between BS $b$ and UE $\psi_b^d(t)$, and the channel between UE $\psi_i^u(t)$ and UE $\psi_b^d(t)$, respectively. It includes path loss, small-scale fading, and shadowing. Further, $x_{\{{}\}}(t)$ is used to denote the complex data symbol transmitted by different nodes. The self-interference channel at BS $b$ is denoted by $h'_{bb}$, which includes the cancellation. We model the transmitted symbols as independent random variables with zero mean and variance $\mathbb{E}\{|x_{\{\}}(t)|^2\} \overset{\Delta}{=} p_{\{\}}(t) \geq 0$. The notation $n_{\psi_b^d(t)}$ and $n_{b}$ denote the additive noise at UE ${\psi_b^d(t)}$ and BS $b$, treated as complex Gaussian random variables with variances $\mathcal{N}_{\psi_b^d(t)}/2$ and $\mathcal{N}_{b}/2$, respectively. 

The signal to interference plus noise (SINR) for downlink UE $\psi_b^d(t)$ and uplink UE $\psi_b^u(t)$ are given by, respectively, 
\begin{equation}\label{SM3}
\begin{split}
&\text{SINR}_{b,\psi_b^d(t)}^d = \\
 &\frac{p_b(t)~G_{b,\psi_b^d(t)}}{ \sum\limits_{i \in \Pi\setminus b} p_i(t)~G_{i,\psi_b^d(t)} + \sum\limits_{i \in \Pi} p_{\psi_i^u(t)}(t)~G_{\psi_i^u(t),\psi_b^d(t)} + \mathcal{N}_{\psi_b^d(t)} },
\end{split}
\end{equation}
\begin{equation}\label{SM4}
\begin{split}
&\text{SINR}_{b,\psi_b^u(t)}^u = \\
&\frac{p_{\psi_b^u(t)}(t)~G_{\psi_b^u(t),b}}{\sum\limits_{i \in \Pi \setminus b} p_{\psi_i^u(t)}(t)~G_{\psi_i^u(t),b} + \sum\limits_{i\in \Pi\setminus b} p_i(t)~G_{i,b} +  p_b(t) \gamma + \mathcal{N}_{b}}.
\end{split}
\end{equation}

In the above equations, $G_{m,n} = |h_{m,n}|^2~\forall m,n$. The residual self-interference is modeled as Gaussian noise, the power of which equals the difference between the transmit power of the BS and the assumed amount of self-interference cancellation. In (\ref{SM4}), $\gamma$ denotes the self interference cancellation level at the BS. The corresponding achievable information rate in \textit{bits/s/Hz} is given by the following Shannon formulas, 
\begin{equation}\label{SM5}
R_{b, \psi_b^d(t)}^d(t) =  \mathrm{log_2}(1+{\text{SINR}}_{b,\psi_b^d(t)}^d),
\end{equation}
\begin{equation}\label{SM6}
R_{b,\psi_b^u(t)}^u(t) = \mathrm{log_2}(1+{\text{SINR}}_{b,\psi_b^u(t)}^u).
\end{equation}

\subsection{Problem Formulation}\label{subsec:SMPFS2}
We consider a system in which there is coordination among the cells. The objective of the coordinated cells is to maximize the system throughput while maintaining a level of fairness among the UEs. We consider a proportional fairness based allocation, which is achieved by maximizing the logarithmic sum of the average rates of all the UEs~\cite{Stoylar05}\cite{PF_proof}. In the FD system both uplink and downlink transmissions need to be considered simultaneously. The objective at timeslot $t$ is defined as
\begin{equation}\label{SM7}
\begin{split}
    & \text{Maximize} \sum\limits_{b \in \Pi} \sum\limits_{k \in \mathcal{K}_b} \left[ \mathrm{log}(\overline{R^d_{b,k}}(t)) + \mathrm{log}(\overline{R^u_{b,k}}(t))\right]  \\
    &\mbox{subject to:} \\
    & \ \ \ \ \ \ \ \  0 \le p_{b}(t) \le p_{\mathrm{max}}^{d},\\
    & \ \ \ \ \ \ \ \ 0 \le p_{k}(t) \le p_{\mathrm{max}}^{u},\\
    & \ \ \ \ \ \ \ \ R^d_{b,k}(t) . R^u_{b,k}(t) = 0, \forall k \in \mathcal{K}_b, \forall b \in \Pi,
\end{split}
\end{equation} 
where $\overline{R^d_{b,k}}(t)$, $\overline{R^u_{b,k}}(t)$ are the average achieved downlink and uplink rates of UE $k$ in cell $b$, denoted as $\mathrm{UE}_{b,k}$, until timeslot $t$, respectively. The first two constraints in (\ref{SM7}) are for the transmit powers of the BSs and UEs in each cell, in which $p_{\mathrm{max}}^{d}$ and $p_{\mathrm{max}}^{u}$ are the maximum powers that can be used in downlink and uplink transmission directions, respectively. The third constraint in (\ref{SM7}) captures the HD nature of the UEs, where $R^d_{b,k}(t) $ and $R^u_{b,k}(t)$ are the instantaneous downlink and uplink rates in timeslot $t$, respectively, of $\mathrm{UE}_{b,k}$ as defined in (\ref{SM5}) and (\ref{SM6}). The average achieved data rate, for example, in downlink, $\overline{R^d_{b,k}}(t)$ is updated iteratively based on the scheduling decision in timeslot $t$, that is, 
\begin{equation}\label{SM8}
\begin{split}
	&\overline{R^d_{b,k}}(t) = \\
	&\left\{ \, 
		\begin{IEEEeqnarraybox}[] [c] {l?s}
			\IEEEstrut
			 \beta\overline{R^d_{b,k}}(t-1) + (1-\beta)R^d_{b,k}(t),~&{if $\psi_b^d(t) = \mathrm{UE}_{b,k}$}, \\
			 \beta\overline{R^d_{b,k}}(t-1),  &  otherwise.
			\IEEEstrut
		\end{IEEEeqnarraybox}
	\right.
	\end{split}
 \end{equation}
where $0<\beta<1$ is a constant weighting factor, which is used to calculate the length of the sliding time window, i.e., $1/(1-\beta)$, over which the average rate is computed for each frame, with its value generally chosen close to one, e.g., 0.99~\cite{Stoylar05, jalali2000data}. The average achieved uplink rate of $\mathrm{UE}_{b,k}$, $\overline{R^u_{b,k}}(t)$ can be similarly defined.

The goal of the coordinated cells is to determine 1) \textit{the set of co-channel UEs scheduled at the same time,} and 2) \textit{the power allocation for the scheduled UEs}, so that the overall utility defined in (\ref{SM7}) can be maximized. 

Assume that ${\mathrm{S}}_b = \{i,j: i \neq j \} \in \mathcal{K}'_b \times \mathcal{K}'_b$ denotes all the possible combinations of choosing two UEs, i.e., one in downlink and one in uplink in cell $b$, where $\mathcal{K}'_b =  \mathcal{K}_b\cup\{\varnothing\}$. $\varnothing$ is used to include the case of no UE selection in a direction. $\boldsymbol{\mathrm{S}} = {\mathrm{S}_1} \times {\mathrm{S}_2} \cdots \times {\mathrm{S}_M}$ is the selection of all UE's in the network. Further, let ${\mathrm{Q}}^{\mathrm{S}_b} = \{p_b,p_j\}, p_b \leq p_{\mathrm{max}}^{d}, p_j \leq p_{\mathrm{max}}^{u}$, denote all possible combination of power levels in the downlink and uplink in ${\mathrm{S}}_b$, and $\boldsymbol{\mathrm{Q}}^{\boldsymbol{\mathrm{S}}}$ =  $[{\mathrm{Q}}^{\mathrm{S}_1},\cdots, {\mathrm{Q}}^{\mathrm{S}_M}] $.

Assume $\boldsymbol{\Psi}(t) \subset \boldsymbol{\mathrm{S}}$ denotes the set of chosen UEs in both downlink and uplink directions in timeslot $t$, i.e., $\boldsymbol{\Psi}(t)$ = $[\{\psi_1^d(t),\psi_1^u(t)\},\cdots,\allowbreak \{\psi_M^d(t),\psi_M^u(t)\} ]$, where $\psi_i^d(t) =\varnothing$ ($\psi_i^u(t) = \varnothing$) indicates no UE scheduled for the downlink (uplink) in cell $i$. This could be the result of no downlink (uplink) demand in cell $i$, in the current time slot $t$; or, as discussed in the next section, it could also be because scheduling any downlink (uplink) transmission in cell $i$, in timeslot $t$ will generate strong interference to the other UEs, lowering the total network utility. So, in each timeslot, each cell will select at most one UE in the downlink and at most one UE in the uplink direction. Assume that $\boldsymbol{\mathcal{P}}(t)$ = $[ \{p_1(t),p_{\psi_1^u(t)}(t)\},\cdots,\{p_M(t),p_{\psi_M^u(t)}(t)\} ]$, where $\boldsymbol{\mathcal{P}}(t) \subset \boldsymbol{\text{Q}}^{\boldsymbol{\Psi}(t)}$ contains the power allocation for the selected UE combination, $\boldsymbol{\Psi}(t)$, in timeslot $t$. 

Using (\ref{SM8}), the objective function in (\ref{SM7}) can be expressed as
\begin{equation}\label{SM9}
\begin{split}
 & \textstyle \sum\limits_{b \in \Pi} \sum\limits_{k \in \mathcal{K}_b} \Big[ \mathrm{log}(\overline{R^d_{b,k}}(t)) + \mathrm{log}(\overline{R^u_{b,k}}(t)) \Big] = \\
 &\sum\limits_{b \in \Pi}  \bigg[ \Big\{ \mathrm{log}\big(\beta\overline{R^d_{b,\psi_b^d(t)}}(t-1) + (1-\beta)R^d_{b,\psi_b^d(t)}(t)\big) - \\
 & \textstyle \mathrm{log}\big(\beta\overline{R^d_{b,\psi_b^d(t)}}(t-1)\big) \Big\} + \Big\{ \mathrm{log}\big(\beta\overline{R^u_{b,\psi_b^u(t)}}(t-1) + \\
 &(1-\beta)R^u_{b,\psi_b^u(t)}(t)\big) - \mathrm{log}\big(\beta\overline{R^u_{b,\psi_b^u(t)}}(t-1)\big) \Big\} \bigg] + A,
 \end{split}
\end{equation}
where $A$ is independent from the decision made at timeslot $t$, and is given by
\begin{equation}\label{SM10}
A =\sum\limits_{b \in \Pi} \sum\limits_{k \in \mathcal{K}_b} \left[ \mathrm{log}(\beta\overline{R^d_{b,k}}(t-1)) + \mathrm{log}(\beta\overline{R^u_{b,k}}(t-1))\right].
\end{equation}

In equation (\ref{SM9}), let us denote the first term in the summation as $\chi^d_{b,\psi_b^d(t)}(t)$,
\begin{equation}\label{SM11}
\begin{split}
\chi^d_{b,\psi_b^d(t)}(t) = &\mathrm{log}(\beta\overline{R^d_{b,\psi_b^d(t)}}(t-1) + \\
&(1-\beta)R^d_{b,\psi_b^d(t)}(t)) - \mathrm{log}(\beta\overline{R^d_{b,\psi_b^d(t)}}(t-1)),
 \end{split}
\end{equation}
which can be further written as,
\begin{equation}\label{SM11_1}
\chi^d_{b,\psi_b^d(t)}(t) = \mathrm{log}\left(1 +  w_{b,\psi_b^d(t)}(t)~R^d_{b,\psi_b^d(t)}(t)\right),
\end{equation}
where
\begin{equation}\label{SM11_2}
w_{b,\psi_b^d(t)}(t) =  \frac{(1-\beta)}{\beta\overline{R^d_{b,\psi_b^d(t)}}(t-1)}.
\end{equation}

Similarly, let us write the second term in (\ref{SM9}) as $\chi^u_{b,\psi_b^u(t)}(t)$,
\begin{equation}\label{SM12}
\chi^u_{b,\psi_b^u(t)}(t) = \mathrm{log}\left(1 + w_{b,\psi_b^u(t)}(t)~R^u_{b,\psi_b^u(t)}(t)\right),
\end{equation}
where
\begin{equation}\label{SM12_1}
w_{b,\psi_b^u(t)}(t) =  \frac{(1-\beta)}{\beta\overline{R^u_{b,\psi_b^u(t)}}(t-1)}.
\end{equation}

In the above equations, note that, if $\psi_b^d(t) = 0 \ (\psi_b^u(t) = 0)$, then $\chi^d_{b,\psi_b^d(t)}(t) = 0  \ (\chi^u_{b,\psi_b^u(t)}(t) = 0)$. The overall utility of a cell (e.g., cell $b$) is defined as
\begin{equation}\label{SM13}
\Phi_{b,\{\psi_b^d(t),\psi_b^u(t)\}}(t) = \chi^d_{b,\psi_b^d(t)}(t) + \chi^u_{b,\psi_b^u(t)}(t).
\end{equation}

Then the optimization problem in (\ref{SM7}) can be equivalently expressed as
\begin{equation}\label{SM14}
{\boldsymbol{\Psi}(t), \boldsymbol{\mathcal{P}}(t)} = \argmax_{\boldsymbol{\mathrm{S}}, \boldsymbol{\mathrm{Q}}^{\boldsymbol{\mathrm{S}}}} \sum\limits_{b \in \Pi} \Phi_{b, \mathrm{S}_b }(t).
\end{equation}

The above problem is a non-linear non-convex combinatorial optimization problem and the optimal solution may not be feasible to compute in practice. Moreover, the above problem is a mixed discrete (UE selection) and continuous (power allocation) optimization. Although the problem can be optimally solved via exhaustive search, the complexity of this method increases exponentially as the number of cells/UEs increases. We will next provide a suboptimal solution of the the above problem which jointly determines the UE selection and power allocation in a distributed manner. 

\section{Channel Estimation in Full Duplex Multi-Cell Networks}\label{sec:CEP}
As discussed in Section~{\ref{sec:intro}}, in a FD multi-cell scenario, channel state information is essential to maximize FD gains. There are three different types of channels to monitor (I) BS-to-UE or UE-to-BS channels; (II) BS-to-BS channels; and (III) UE-to-UE channels. Since we assume a TDD system in this paper, the channels between any two radios in both directions are reciprocal. Existing 3GPP protocols for HD communications already include mechanisms to monitor type I channels, in which a terminal (UE) needs to estimate the channel with a BS. In 3GPP LTE, cell-specific reference signals are broadcast from the BSs with their physical-layer cell identity. UEs then use the received reference signals to estimate the channels from the BSs and transmit channel state information (CSI) reports to BSs using PUCCH and PUSCH \cite{DahlmanLTE}\cite{3GPP:5}. The same signal can be used at the BS receiver to estimate the channel from its neighboring BSs, i.e., type II channels. The remaining challenge for the FD multi-cell scenario is to estimate UE-to-UE interference, or type III channels, since the inter-UE interference poses a fundamental challenge to exploit FD in a cellular scenario. 

In this paper, we propose to implement neighbor discovery at UEs to find potential UE interferers in its neighborhood. In 3GPP LTE, Sounding Reference Signals (SRS) are used for channel quality estimation at different frequencies in the uplink~\cite{DahlmanLTE}. This uplink SRS can be used by UEs to estimate the channels with other UEs in its neighborhood~\cite{D2DneighborDisSRS}. In LTE, each UE is scheduled on the SRS channel regularly in order for the BS (eNB) to collect information for uplink channel scheduling. All UEs within a cell are informed about the subframes that will be used for SRS. The main challenge in neighbor discovery is to distinguish between different UEs, including neighboring cells' UEs, during SRS transmission. This problem can be solved by allocating different SRS combination sets to neighboring cells as well as different orthogonal combinations to UEs within the cell which are scheduled to transmit simultaneously~\cite{DahlmanLTE}. In addition, this allocation of SRS combinations can be passed to UEs through the downlink shared channel \cite{D2DneighborDisSRS}. There are alternate ways to implement neighbor discovery, such as mechanisms proposed for D2D communications~\cite{D2DneighborDisDMRS, D2DneighborDisFlashLinq}. In this paper, for our scheduling solution we assume that each UE will be able to estimate the channels within its neighborhood, i.e., channels with strong UE interferers, and this information will be transmitted to its BS. The signaling overhead during the transmission of such new UE-to-UE channel information over the air link in analyzed in Section~\ref{sec:PE}.

\section{A Distributed Full Duplex Multi-Cell Resource Allocation (DFDMR)}\label{sec:DFDMR}
In this section we provide a distributed method to solve (\ref{SM14}). As discussed in Section~\ref{sec:intro}, FD throughput gain is available only under certain propagation conditions, distances among nodes in the network, and power levels. This suggests that FD operation should be used opportunistically, that is, with an intelligent scheduler that schedules UEs with appropriate power levels to achieve FD operation when appropriate, and otherwise defaults to HD operation. In each timeslot, the joint UE selection and power allocation problem (\ref{SM14}) is solved in two steps, (1) $\textit  {Intra-cell UE Selection}$: for a given feasible power allocation, this step finds the UE or a pair of UEs in each cell with maximum overall utility, and (2) $\textit {Inter-cell Coordination} $: for the given UE selection, this step derives the powers to be allocated to the selected UEs through inter-cell coordination such that overall utility can be maximized. In the next subsections, we discuss both steps in detail. 

\subsection{Intra-cell UE Selection}\label{sec:DSS1}
In this step, for each timeslot $t$, each BS selects the UE or a pair of UEs to be scheduled. This is a single cell resource allocation problem, which can be solved in multiple ways \cite{DiINFOCOM, Shaocommletter,GMSingleCell}. Given the fact that a small cell does not have many UEs, it is easy to perform resource allocation in a centralized manner at the BS. The BS has knowledge of the channel gains with its all UEs, which is possible through CSI reporting from its UEs \cite{DahlmanLTE}\cite{3GPP:5}. As discussed in Section~\ref{sec:CEP}, we further assume that the BS also knows the channel between all UE pairs and thus the subset of UE pairs with strong mutual interference. The BS will assume no interference between UE pairs for which no information is received, presumably because of a weak SRS signal. 
%Although the distributed methods for this problem can also be easily designed, in which UEs do not need to transmit the UE-to-UE channels information for all its UEs to its BS. Since more delay incurs to implement the procedure in a distributed manner, so choosing a centralized approach for the single small cell is a better idea where the complexity is not significant. 

In this step, each BS $b \in \Pi$, for the given feasible power allocation, finds the UEs which provide the maximum utility defined in (\ref{SM13}),
\begin{equation}\label{DS1}
	\{\psi_b^d(t), \psi_b^u(t) \} =\argmax_{\substack{{\mathrm{S}}_b}}  \Phi_{b,\boldsymbol{\text{S}}_b}(t).
\end{equation} 

Please note that at this stage, there is no inter-cell information available, so in the above equation, the instantaneous rate of a UE does not take any inter-cell interference into account. Thus, for the cell $b$, instead of (\ref{SM3}) and (\ref{SM4}), the $\text{SINRs}$ at downlink UE $i$ and uplink UE $j$ are calculated as 
\begin{equation}\label{DS2}
\text{SINR}_{b,i}^d = \frac{p_b(t)~G_{b,i}}{ p_{j}(t)~\widetilde{G}_{j,i} + \mathcal{N}_{i} }, ~~\text{SINR}_{b,j}^u = \frac{p_{j}(t)~G_{j,b}}{ p_b(t) \gamma + \mathcal{N}_{b}},
\end{equation}
where $\widetilde{G}_{ji}$ denotes the channel gain estimation between UE $j$ and UE $i$ measured by UE $i$. If UE $i$ does not hear a strong signal from UE $j$, this means UE $i$ did not measure and send the channel estimation information for UE $j$ to the BS. In that case $\widetilde{G}_{ji}$ will be neglected during this scheduling decision. The problem (\ref{DS1}) can be solved simply by the exhaustive search method. The BS initially assumes the maximum power allocation for each UE in both directions, and then calculates the aggregate utility for each possible combination of UEs and finds the utility maximizing UE or UEs. Since each cell performs this step independently, the computation complexity of this step increases only in a quadratic manner with the number of UEs, i.e., $O(n^2)$, which should not be a problem given that a small cell typically supports a small number of UEs. After this step, each cell has a downlink UE, or an uplink UE, or both to schedule in timeslot $t$. Once the UE selection is done, the next step is inter-cell coordination, described next, in which the power levels of the selected UEs are updated such that the aggregate utility of all the UEs, as given in (\ref{SM14}), can be maximized.

\subsection{Inter-cell Coordination}\label{sec:DSS2}
This step is used to take the effect of inter-cell interference into account. In this step, the transmit power levels of all the selected UEs are updated such that the mutual interference can be mitigated and the overall utility of the system can be maximized. The objective function of this problem can be written as,
\begin{equation}\label{DS5}
     \boldsymbol{\mathcal{P}}(t) = \argmax_{\boldsymbol{\mathrm{Q}}^{\boldsymbol{\Psi}(t)}} \sum\limits_{b \in \Pi} \Phi_{b,\{\psi_b^d(t),\psi_b^u(t)\}}(t).\\
 \end{equation}

Each of the BSs solves the above problem independently and derives its optimum powers. The utilities of the other BSs are estimated based on the information received from neighboring BSs. 
The detailed procedure is given below. This procedure is completed in multiple iterations. It is assumed that the information between the BSs is exchanged over the X2 interface \cite{3GPP:4}. Note that this procedure is applied at each timeslot, but for the sake of simplifying the notation, we omit the term $t$ in this section. 

1) \textit{Initialization}:  Intra-cell UE selection determines the UEs to be scheduled, i.e., $\psi_b^d$, $\psi_b^u$ in cell $b\in\Pi$. At this initial step, each BS $b \in\Pi$ broadcasts a message vector containing the information of weights $(w_{b,\psi_b^d},w_{b,\psi_b^u})$, UE IDs $(id(\psi_b^d), id(\psi_b^u))$, and the channel gains $(G_{b,\psi_b^d}, G_{\psi_b^u,b})$ with its own BS for the selected UEs. In addition to this information, the channel gains of the selected UEs with other BSs are also sent to the corresponding BSs. For example channel gains with BS $j$, i.e., $(G_{j,\psi_b^d}, G_{\psi_b^u,j})$ are sent to the BS $j$. This information is only sent once at the initialization step. Here, we use UE IDs corresponding to the value of SRS combination allocated to a UE. The UE IDs of other cells's UEs will be used at a BS to identify and match the UE-to-UE channels estimations measured by its own cells's UEs. These IDs can be created locally at each BS by matching UEs to the allocation of SRS combinations. In addition to the above information, after getting UE IDs information, each BS also sends some required UE-to-UE channel information as described further in this section. 
  
2) \textit{Power Update}:  After the initial information exchange, each iteration $(n \geq 1)$ has two steps: 
	
\textit{First Step}: Each BS calculates the total received uplink and downlink interference based on the information received during initialization and in the previous iteration $(n-1)$. For example, in BS $b \in \Pi$, the estimated interference in downlink and uplink are given, respectively, by 
 \begin{equation}\label{DS10}
 \small{
     \boldsymbol{\text{I}}_{\psi_b^d}^{(n-1)}  =  \mathcal{N}_{\psi_b^d} + \sum\limits_{i \in \Pi\setminus b} p_i^{(n-1)}~{{G}_{i,\psi_b^d}} + \sum\limits_{i \in \Pi} p_{\psi_i^u}^{(n-1)}~\widetilde{G}_{\psi_i^u,\psi_b^d} ,
     }
\end{equation} 
\begin{equation}\label{DS11}
\small{
   \boldsymbol{\text{I}}_b^{(n-1)}  = p_b^{(n-1)} \gamma + \mathcal{N}_{b} +  \sum\limits_{i \in \Pi \setminus b} p_{\psi_i^u}^{(n-1)}~{G}_{\psi_i^u,b} + \sum\limits_{i\in \Pi\setminus b} p_i^{(n-1)}~{G}_{i,b},   
   }
\end{equation}  
where $p^{(n-1)}_{\{\}}$ is the power values derived in the previous iteration as discussed in the next step; $\widetilde{G}_{\psi_i^u,\psi_b^d}$ is the channel measured by UE $\psi_b^d$ with $\psi_i^u$ of cell $i$ as discussed in Section~\ref{sec:CEP}. The UE IDs information exchanged during initialization is used during this process.  
 
At the end of this step, the value of the estimated interference is broadcast by each BS to its neighbors. 

\textit{Second Step}: Each BS updates its transmit powers to maximize the aggregate utility sum ($\ref{DS5}$), given the power levels of other transmitters at the previous iteration, and the interference information received in the first step. 

At each BS $b \in \Pi$,
\begin{equation}\label{DS12}
\{ p_b^{(n)}, p_{\psi_b^u}^{(n)}\}  =  \argmax_{\{x,y\} \in \mathrm{Q}^{\{\psi_b^d, \psi_b^u\}}} \sum\limits_{j \in \Pi} \widetilde{\Phi}^{b,(n-1)}_{j,\{\psi_j^d,\psi_j^u\}},
\end{equation}
where $\widetilde{\Phi}^{b,(n-1)}_{\{\}}$ is the estimated value of the overall utility calculated at BS $b$. It can be written as
\begin{equation}\label{DS13}
\begin{split}
 &\{ p_b^{(n)}, p_{\psi_b^u}^{(n)}\}  = \\
 &\argmax_{\{x,y\} \in \mathrm{Q}^{\{\psi_b^d, \psi_b^u\}}}  \sum\limits_{j \in \Pi} \bigg[ \mathrm{log}\Big( 1 +  w_{j,\psi_j^d}~{\mathrm{log}}_2 \big( 1 + {\text{SINR}_{j,\psi_j^d}^{b,(n-1)}}\big) \Big)  \\
 &+ \mathrm{log}\Big( 1 +  w_{j,\psi_j^u}~{\mathrm{log}}_2 \big( 1 + {\text{SINR}_{j,\psi_j^u}^{b,(n-1)}}\big) \Big) \bigg],
\end{split}
\end{equation} 
where,
\begin{equation}\label{DS14}
\small{
\begin{split}
	&\text{SINR}_{j,\psi_j^d}^{b,(n-1)} = \\
	&\left \{ \,
		\begin{IEEEeqnarraybox}[] [c] {l?s}
			\IEEEstrut
			\frac{x~{G}_{b,\psi_b^d} }{ \boldsymbol{\text{I}}_{\psi_b^d}^{(n-1)} + (y- p_{\psi_b^u}^{(n-1)}) \widetilde{G}_{\psi_b^u, \psi_b^d}} &  $j = b$, \\
			\frac{p_{j}^{(n-1)}~{G}_{j,\psi_j^d} }{ \boldsymbol{\text{I}}_{\psi_j^d}^{(n-1)} +  (x- p_b^{(n-1)}) {G}_{b,\psi_j^d} + (y- p_{\psi_b^u}^{(n-1)}) \widetilde{G}_{\psi_b^u,\psi_j^d}} & $ j \neq b$, 
			\IEEEstrut
		\end{IEEEeqnarraybox}  
		\right. 	
\end{split}}
\end{equation}

\begin{equation}\label{DS15}
\small{
\begin{split}
	&{\text{SINR}_{j,\psi_j^u}^{b,(n-1)}} = \\
	&\left \{ \,
		\begin{IEEEeqnarraybox}[] [c] {l?s}
			\IEEEstrut
			\frac{y~{G}_{\psi_b^u,b} }{ \boldsymbol{\text{I}}_{\psi_b^u}^{(n-1)} + (x- p_{\psi_b^d}^{(n-1)}) \gamma} &  $j = b$, \\[-5pt]
			\frac{p_{\psi_j^u}^{(n-1)}~{G}_{\psi_j^u,j} }{ \boldsymbol{\text{I}}_{\psi_j^u}^{(n-1)} +  (x- p_b^{(n-1)}) {G}_{b,j} + (y- p_{\psi_b^u}^{(n-1)}) {G}_{\psi_b^u,j}} & $ j \neq b$, 
			\IEEEstrut
		\end{IEEEeqnarraybox}  
	\right. 
\end{split}}
\end{equation}
 
Note that in (\ref{DS14}), the channel $\widetilde{G}_{\psi_b^u,\psi_j^d}$ is measured at $\psi_j^d$ in cell $j$ as described in Section~\ref{sec:CEP}. This information is sent by BS $j$ to BS $b$ after receiving UE IDs of the selected UEs during the initialization process.  

We use GP~\cite{boyd2007tutorial, chiang2007power} to get a near-optimal solution of this nonlinear nonconvex optimization~(\ref{DS13}). GP cannot be applied directly to the objective function given in (\ref{DS13}), so we first convert our objective function into a weighted sum rate maximization using the following approximation. In (\ref{DS13}), for the weight terms, let us consider $w_{j,\psi_j^d}$, which is given by (\ref{SM11_2}). Since we set $\beta$ very close to one, and moreover, if we assume that the value of the instantaneous rate, $R^d_{j,\psi_j^d}$, will be of the same order as the average rate, $\overline{R^d_{j,\psi_j^d}}$, then the term $\frac{(1-\beta)R^d_{j,\psi_j^d}}{\beta\overline{R^d_{j,\psi_j^d(t)}}}$ will be close to zero. So, by using $ln(1+x) \approx x$ for $x$ close to zero, (\ref{DS13}) can be approximated by
\begin{equation}\label{DS16}
\begin{split}
 &\{ p_b^{(n)}, p_{\psi_b^u}^{(n)}\}  = \\
 &\argmax_{\{x,y\} \in \mathrm{Q}^{\{\psi_b^d, \psi_b^u\}}}  \sum\limits_{j \in \Pi}  \Big( w_{j,\psi_j^d}~{\mathrm{log}}_2 \big( 1 + {\text{SINR}_{j,\psi_j^d}^{b,(n-1)}}\big)  \\
 & + w_{j,\psi_j^u}~{\mathrm{log}}_2 \big( 1 + {\text{SINR}_{j,\psi_j^u}^{b,(n-1)}}\big) \Big),
\end{split}
\end{equation}
 
Please note that both $x$ and $y$ in $\mathrm{Q}^{\{\psi_b^d, \psi_b^u\}}$ have inbuilt maximum power constraint given in (\ref{SM7}). The problem (\ref{DS16}) can be further written as
\begin{equation}\label{DS17}
\footnotesize{
\begin{split}
    & \argmin_{\{x,y\}}  {\prod_{j = 1}^{M} \Bigg( \bigg( \frac{1}{1+{\text{SINR}}^{b,(n-1)}_{j,\psi_j^d}}\bigg)^{w_{j,\psi_j^d}}. \bigg( \frac{1}{1+{\text{SINR}}^{b,(n-1)}_{j,\psi_j^u}}\bigg)^{w_{j,\psi_j^u}} \Bigg)}\\
    &\mbox{subject to:} \\
    & \ \ \ \ \ \ \ \ 0 \le \frac{x}{p_{\mathrm{max}}^{d}} \le 1, ~~ 0 \le \frac{y}{p_{\mathrm{max}}^{u}} \le 1
\end{split}}
\end{equation} 

In general, to apply GP, the optimization problem should be in GP standard form~\cite{boyd2007tutorial, chiang2007power}. In the GP standard form, the objective function is a minimization of a $\textit {posynomial}$\footnote{ A monomial is a function $f:\mathbf{R}_{++}^n \rightarrow \mathbf{R}: g(p) = d p_1^{a^{(1)}}p_2^{a^{(2)}}\cdots p_n^{a^{(n)}}$, where $d \geq 0$ and $a^{(k)} \in \mathbf{R}, k = 1,2,\cdots,n.$ A posynomial is a sum of monomials, $f(p) = \sum_{j=1}^J d_j p_1^{a_j^{(1)}} p_2^{a_j^{(2)}} \cdots p_n^{a_j^{(n)}}$. } function; the inequalities and equalities in the constraint set are a posynomial upper bound inequality and $\textit {monomial}$ equality, respectively. 

In our case, in (\ref{DS17}), constraints are monomials (hence posynomials), but the objective function is a ratio of posynomials, as shown in  (\ref{DS18}). Hence, (\ref{DS17}) is not a GP in standard form, because posynomials are closed under multiplication and addition, but not in division.

\begin{figure*}[!t]
% ensure that we have normalsize text
\normalsize
% Store the current equation number.
\setcounter{mytempeqncnt}{\value{equation}}
% Set the equation number to one less than the one
% desired for the first equation here.
% The value here will have to changed if equations
% are added or removed prior to the place these
% equations are referenced in the main text.
\setcounter{equation}{28}
\begin{equation}\label{DS18}
\begin{split}
& \textstyle {\prod_{j = 1}^{M} \Bigg( \bigg( \frac{1}{1+{\text{SINR}}^{b,(n-1)}_{j,\psi_j^d}}\bigg)^{w_{j,\psi_j^d}}. \bigg( \frac{1}{1+{\text{SINR}}^{b,(n-1)}_{j,\psi_j^u}}\bigg)^{w_{j,\psi_j^u}} \Bigg)} \\
 &= \textstyle { \bigg( \frac{\boldsymbol{\text{I}}_{\psi_b^d}^{(n-1)} + (y- p_{\psi_b^u}^{(n-1)}) \widetilde{G}_{\psi_b^u, \psi_b^d}}{\boldsymbol{\text{I}}_{\psi_b^d}^{(n-1)} + (y- p_{\psi_b^u}^{(n-1)}) \widetilde{G}_{\psi_b^u, \psi_b^d} + x~{G}_{b,\psi_b^d} }\bigg)^{w_{b,\psi_b^d(t)}}}. \scriptstyle { \bigg( \frac{\boldsymbol{\text{I}}_{\psi_b^u}^{(n-1)} + (x- p_{\psi_b^d}^{(n-1)}) \gamma}{\boldsymbol{\text{I}}_{\psi_b^u}^{(n-1)} + (x- p_{\psi_b^d}^{(n-1)}) \gamma + y~{G}_{\psi_b^u,b}}\bigg)^{w_{b,\psi_b^d(t)}}}. \\
&\textstyle { \prod_{j=1, j \neq b}^M \Bigg( \bigg( \frac{\mathcal{C}_{\psi_j^d}^{b,(n-1)} +  x {G}_{b,\psi_j^d} + y \widetilde{G}_{\psi_b^u,\psi_j^d}}{\mathcal{C}_{\psi_j^d}^{b,(n-1)} +  x {G}_{b,\psi_j^d} + y \widetilde{G}_{\psi_b^u,\psi_j^d} + p_{j}^{(n-1)}{G}_{j,\psi_j^d} }\bigg)^{w_{j,\psi_j^d(t)}}}. \scriptstyle {  \bigg( \frac{\mathcal{C}_{\psi_j^u}^{b,(n-1)} + x {G}_{b,j} + y {G}_{\psi_b^u,j}}{\mathcal{C}_{\psi_j^u}^{b,(n-1)} + x {G}_{b,j} + y {G}_{\psi_b^u,j} +p_{\psi_j^u}^{(n-1)}{G}_{\psi_j^u,j}  }\bigg)^{w_{j,\psi_j^u(t)}} \Bigg)} \\
&\text{where} \\
&\mathcal{C}_{\psi_j^d}^{b,(n-1)} = \boldsymbol{\text{I}}_{\psi_j^d}^{(n-1)} - p_b^{(n-1)} {G}_{b,\psi_j^d} - p_{\psi_b^u}^{(n-1)} \widetilde{G}_{\psi_b^u,\psi_j^d}, \\
&\mathcal{C}_{\psi_j^u}^{b,(n-1)} = \boldsymbol{\text{I}}_{\psi_j^u}^{(n-1)} - p_b^{(n-1)} {G}_{b,j} - p_{\psi_b^u}^{(n-1)} {G}_{\psi_b^u,j}
\end{split}
\end{equation}
% Restore the current equation number.
\setcounter{equation}{29}
% IEEE uses as a separator
\hrulefill
% The spacer can be tweaked to stop underfull vboxes.
\vspace*{4pt}
\end{figure*}

According to~\cite{chiang2007power}, (\ref{DS17}) is a signomial programming (SP) problem. In~\cite{chiang2007power}, an iterative procedure is given, in which (\ref{DS17}) is solved by constructing a series of GPs, each of which can easily be solved. In each iteration$\footnote{Please note that this iterative procedure to solve GP is an inner procedure of the main iterative procedure of the distributed \textit{Power Update} step.}$ of the series, the GP is constructed by approximating the denominator posynomial (\ref{DS18}) by a monomial, then using the arithmetic-geometric mean inequality and the value of $\{x,y\}$ from the previous iteration. The series is initialized by any feasible $\{x,y\}$, and the iteration is terminated at the $s_{th}$ loop if $||{x}_s - x_{s-1}|| < \epsilon $, and $||{y}_s - y_{s-1}|| < \epsilon $, where $\epsilon$ is the error tolerance. This procedure is provably convergent, and empirically almost always computes the optimal power allocation~\cite{chiang2007power}.  

The new derived values are broadcast by each BS to its neighboring BSs. Then the same procedure is applied starting from the \textit{Power Update} step (step 2) until the termination condition described below is reached .
 
3) \textit{Termination}: The procedure ends when either a maximum number of iterations is reached or a terminating solution is obtained. For the UE selection $\boldsymbol{\Psi}$ given by $\textit{Intra-Cell UE Selection}$, a power allocation $\boldsymbol{\mathcal{P}} \in \boldsymbol{\mathrm{Q}}^{\boldsymbol{\Psi}}$ will be a terminating solution if changing the power level of any single transmitter cannot improve the aggregate utility sum, given the power levels of all other transmitters. It was observed in the simulation results that with the above power update rule, the termination condition is achieved in a few iterations. 

\section{A Centralized Full Duplex Multi-Cell Resource Allocation (CFDMR)}\label{sec:CFDMR}
In this section, to evaluate the performance of our proposed distributed approach against a centralized approach, we describe a centralized solution to solve the problem (\ref{SM14}). We assume a centralized scheduler that has access to global information, i.e., channel state information, power, etc., and jointly derives the UE selection and power allocation for all the cells simultaneously. The results generated using this scheduler can be viewed as an upper bound on system performance. In this setting, as in the decentralized problem, the joint problem of UE selection and power allocation (\ref{SM14}) is solved in two steps, (1) \textit {Greedy UE Selection}, and (2) \textit {Centralized Power Allocation}. 

\subsection{Greedy UE Selection }\label{sec:GUS}
In each timeslot $t$, for a given feasible power allocation, the centralized scheduler finds a UE or a pair of UEs in each cell  to transmit, which is given as
\begin{equation}\label{CS1}
 \boldsymbol{\Psi}(t) =\argmax_{\boldsymbol{\mathrm{S}}}  \sum_{b=1}^M \Phi_{b,\{\psi_b^d(t),\psi_b^u(t)\}}(t)   
\end{equation}

In traditional HD systems, finding the optimal set of UEs is very different in the downlink and uplink direction. In the literature, the problem above is solved optimally in the downlink direction~\cite{venturino2009coordinated, yu2010joint, kiani2007maximizing}, where the interferers are the fixed BSs in the neighboring cells, assuming a synchronized HD multi-cell system. It is easy to estimate the channel gains between each UE with the neighboring BSs. Thus, interference from the neighboring cells can be calculated without knowing the actual scheduling decision (UE selection) of the neighboring cells. In this situation, a centralized scheduler can calculate the instantaneous rate and the utility of each UE in each cell, and make the UE selection decision for each cell optimally. In uplink scheduling, for the given power allocation, interference from the neighboring cell cannot be calculated until the actual scheduling decision of the neighboring cell is known, because in this case, a UE in the neighboring cell generates the interference. This also applies to the FD system, where interference from the neighboring cell could be from a UE or the BS or both. 

To solve this problem, we use a heuristic greedy method similar to~\cite{SanjayCISS13,koutsopoulos2006cross}. In this method, the centralized greedy algorithm runs over a random order of all the cells, and selects UEs in each cell one by one. For each cell, the UE or a pair of UEs are selected with maximum utility gain, where the utility gain is the difference between the gain in the marginal utility of the chosen UE or UEs and the loss in the marginal utility of selected UEs in other cells due to new interference generated from the the cell being considered. Moreover, for the UEs in the cell being considered, interference from only the cells for which decision has been made is considered. Since this is the same method as the one given in \cite{SanjayCISS13}, we omit the details of this algorithm in this paper. The complete algorithm can be found in \cite{SanjayCentArxiv}. This algorithm gives the UE combination $\boldsymbol{\Psi}(t)$. 

\subsection{Centralized Power Allocation }\label{sec:CPA}
In this step, for the selected UE combination in the previous step, a centralized power allocation process is applied to find the appropriate power levels for all UEs, so that the overall utility can be maximized as described in~(\ref{DS5}). In this case, similar to the Section~\ref{sec:DSS2}, we use GP to solve this nonlinear nonconvex problem, but in a centralized manner. Since we assume the centralized scheduler has access to the to global information, GP is applied once\footnote{In this case also it will be a signomial programming, which will be solved in an iterative procedure by constructing a series of GPs.} at the scheduler to find the optimum power allocation for all the selected UEs, instead of applying it independently at each BS as in the Section~\ref{sec:DSS2}. More details can be found in \cite{SanjayCentArxiv} for the centralized power allocation.

\begin{figure*} 
\centering
\includegraphics[width = 4in] {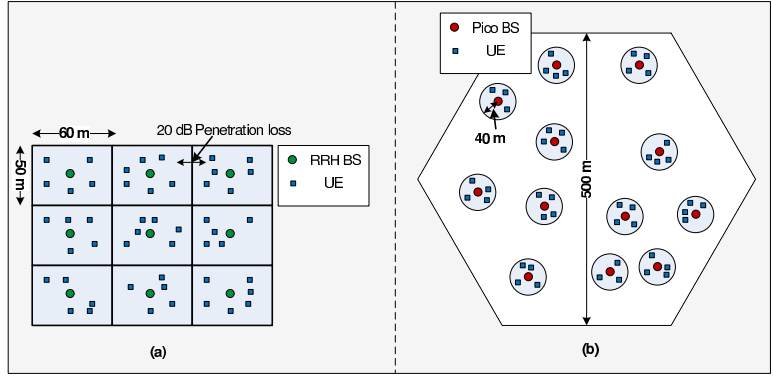}
\caption{(a) An indoor environment with nine RRH Cells, (b) An outdoor environment with twelve picocells.}
\label{fig:fig5}
\end{figure*}

\section{Performance Evaluation}\label{sec:PE}
In this section, we evaluate the performance of the FD system compared to a baseline HD system using the joint UE selection and power allocation presented in Sections~\ref{sec:DFDMR} and~\ref{sec:CFDMR}. To simulate the HD system, we consider both \textit{synchronous} as well as a \textit{dynamic TDD} \cite{3GPP:1} system. In the \textit{synchronous} HD setting, in a given timeslot, all cells schedule either uplink or downlink transmission, and the number of timeslots is divided equally between the uplink and downlink transmission. In \textit{dynamic TDD}, each cell has the flexibility of scheduling its UE in any direction, whichever provides larger utility at the given timeslot. The same distributed and centralized algorithms are also applied to schedule the UEs and to determine the power allocation in these HD systems. For example, for the HD case, (\ref{DS16}), (\ref{DS17}), (\ref{DS18}) will just contain a single term for the corresponding direction instead of two terms.

\subsection{Deployment Scenarios and Simulation Parameters}\label{sec:DSaSP}
We consider both indoor and outdoor deployment scenarios in our simulations. For the indoor environment, a dense multi-cell system with nine indoor Remote Radio Head (RRH)/Hotzone cells, as shown in Figure~\ref{fig:fig5}(a), is considered. The simulation parameters, based on 3GPP simulation recommendations for an RRH cell environment~\cite{3GPP:3}, are described in Table~\ref{tab1}. The path loss for both $LOS$ and $NLOS$ within a cell are given in Table~\ref{tab1}, where the probability of $LOS$ ($P_{LOS}$) is,
\begin{equation}\label{PE1}
\small{
	P_{LOS} = 
	\left \{ \,
		\begin{IEEEeqnarraybox}[] [c] {l?s}
			\IEEEstrut
			1 &  $R \leq 0.018$, \\
			\exp{(-(R-0.018)/0.027)} & $ 0.018 < R < 0.037$, \\
			0.5 & $R \geq 0.037$,
			\IEEEstrut
		\end{IEEEeqnarraybox}  
	\right. 
	}
\end{equation}

In (\ref{PE1}), $R$ is the distance in kilometers. The channel model used between BSs and UEs is also used between UEs, and between BSs for the FD interference calculations, with the justification that BSs do not have a significant height advantage in the small cell indoor scenario considered, and that it is a conservative assumption for the UE-to-UE interference channel. Eight randomly distributed UEs are deployed in each cell. 

To simulate an outdoor multi-cell scenario, the parameters related to path loss, shadowing, and noise figure used in simulations are based on the 3GPP simulation recommendations for outdoor environments~\cite{3GPP:1}, and are described in Table~\ref{tab2}. The probability of LOS for BS-to-BS and BS-to-UE path loss is (R is in kilometers) is
\begin{equation}\label{21}
\begin{split}
P_{LOS} = 0.5 - \text{min}(0.5, 5~\exp(-0.156/R)) + \\
\text{min}(0.5, 5~\exp(-R/0.03)).
\end{split}
\end{equation}

\begin {table*} 
\caption {Simulation parameters for an indoor multi-cell scenario} \label{tab1} 
\begin{center}
    \begin{tabular}{| p{2.6 in} | p{3.4 in} |}
    	\hline
		\textbf {Parameter} & \textbf{Value} \\ \hline
		Noise figure & BS: 8 dB, UE: 9 dB \\ \hline
		Shadowing standard deviation (with no correlation) &  LOS:~ 3 dB NLOS: 4 dB \\  \hline
		Path loss within a cell (dB) (R in kilometers) &  LOS: $89.5 + 16.9~log_{10}(R)$, 
		NLOS: $147.4 + 43.3~log_{10}(R)$ \\ \hline
		Path loss between two cells (R in kilometers) & Max(($131.1 + 42.8~log_{10}(R)),(147.4 + 43.3~log_{10}(R)))$ \\ \hline
		Penetration loss & Due to boundary wall of an RRH cell: 20 dB, Within a cell: 0 dB \\ \hline
    \end{tabular}
\end{center}
\end{table*}

\begin {table*} 
\caption {Simulation parameters for an outdoor multi-cell scenario} \label{tab2} 
\begin{center}
    \begin{tabular}{| p{2.6 in} | p{4 in} |}
    	\hline
		\textbf {Parameter} & \textbf{Value} \\ \hline
		Minimum distance between pico BSs & 40 m \\ \hline
		Radius of a picocell & 40 m \\ \hline
		Noise figure & BS: 13 dB, UE: 9 dB \\ \hline
		Shadowing standard deviation between BS and UE &  LOS:~ 3 dB NLOS: 4 dB \\  \hline
		Shadowing standard deviation between picocells &  6 dB \\  \hline
		BS-to-BS path loss (R in kilometers) & LOS: if $R < 2/3 km, PL(R) = 98.4 + 20~log_{10}(R)$, else $PL(R) = 101.9 + 40~log_{10}(R)$, NLOS: $PL(R) = 169.36 + 40 log_{10}(R)$.  \\ \hline
		BS-to-UE path loss (R in kilometers) & LOS: $PL(R) = 103.8 + 20.9~log_{10}(R),$ NLOS: $PL(R) = 145.4 + 37.5~log_{10}(R)$.  \\ \hline
		UE-to-UE path loss (R in kilometers) & If $R \leq 50 m, PL(R) = 98.45 + 20~log_{10}(R),$ else,  $PL(R) = 175.78 + 40~log_{10}(R)$.  \\ \hline
    \end{tabular}
\end{center}
\end{table*}

For the outdoor environment, we first considered the same dense multi-cell system as shown in Figure~\ref{fig:fig5}(a), assuming no wall(s) between the cells. However, the performance gain of FD operation in such a dense outdoor environments was not substantial due to strong inter-cell interference when no mitigation other than scheduling and power control is applied.
We therefore analyzed the performance of FD operation in a sparse outdoor multi-cell system with twelve randomly dropped picocells, each with ten randomly distributed UEs as shown in Figure~\ref{fig:fig5}(b). This deployment reflects current picocell deployment, which cover local traffic hotspots. As we described in Section~\ref{sec:intro}, since FD operation increases the interference in a network significantly, exploiting FD operation in such an indoor environment or a sparse outdoor environment is more beneficial because of the reduction in inter-cell interference. 

In both indoor and outdoor scenarios, the channel bandwidth is 10 MHz, the maximum BS power is 24 dBm, the maximum UE power is 23 dBm, and the thermal noise density is -174 dBm/Hz. In our simulations, since we use the Shannon equation to measure the data rate, we apply a maximum spectral efficiency of 6 bits/sec/Hz (corresponding to 64-QAM modulation) to match practical systems. BSs and UEs are assumed to be equipped with single omnidirectional antennas. We simulated the system with both full buffer traffic and non-full buffer FTP traffic assumptions. In the next few sections, we present the performance of the FD system with both distributed and centralized scheduling algorithms, and also discuss the convergence and signaling overhead in these methods. In the following sections, we use \emph{FD@x} to represent the FD system with self-interference cancellation of $x$ dB. \emph{FD@Inf} means that there is no self-interference. 

\begin{figure*} 
\centering
\includegraphics[width = 5in] {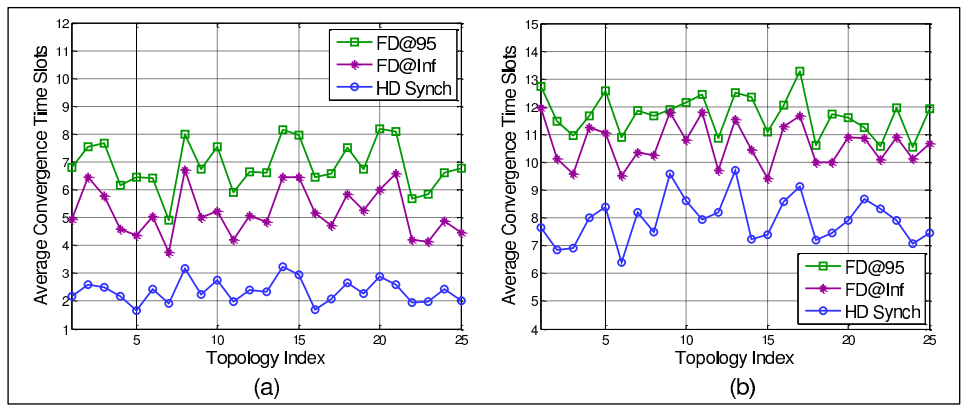}
\caption{(a) Average number of iterations required to converge in different topologies in an (a) indoor multi-cell scenario, (b) outdoor multi-cell scenario.}
\label{fig:convergence_all}
\end{figure*}

\subsection{On the Convergence of DFDMR}
In this section we study the convergence of the distributed scheduling algorithm presented in Section~\ref{sec:DFDMR}. Figure~\ref{fig:convergence_all} shows the average number of iterations required to converge. Figure~\ref{fig:convergence_all}(a) shows the result for the indoor multi-cell case for FD@95, FD@Inf and HD synchronous systems. We calculate the average convergence time taken over different distributions of the UEs, i.e., different topologies. In the FD case, due to higher number of simultaneous transmissions, it takes longer to converge compared to the HD system. Moreover, due to higher interference in FD@95, the scheduler takes longer to converge compared to the FD@Inf system. In the outdoor scenario given in Figure~\ref{fig:fig5}(b), the same trend is observed as shown in the Figure~\ref{fig:convergence_all}(b). In this case, results are obtained with different random drops of pico cells. Due to higher inter-cell interference between a BS and UEs as compared to the indoor scenario, a higher number of iterations are required for the outdoor scenario.

\subsection{Throughput Performance}\label{sec:TP}
With the above simulation settings, in the indoor case, we run our simulation for different UE drops in all cells, each for a thousand timeslots, with the standard wrap around topology, and generate results for both the HD and FD systems. In this section, we simulate the system in which each UE has full-buffer traffic in both directions; the results with the non-full buffer traffic case will be presented in Section~\ref{sec:NFT}.  

\begin{figure*} 
\centering
\includegraphics[width = 5.5in] {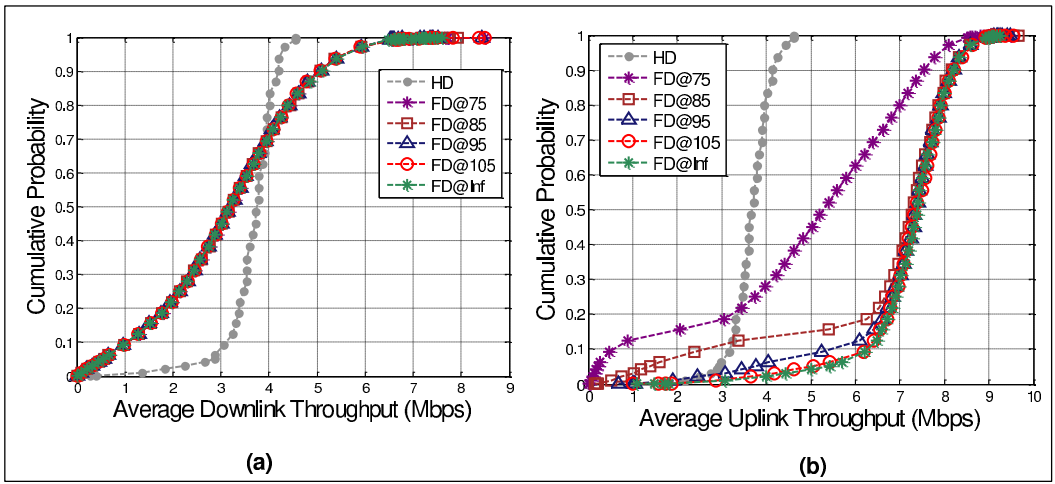}
\caption{Distribution of average data rates for the half duplex system and full duplex system with round-robin scheduler in an indoor multi-cell scenario.}
\label{fig:rr_cdf}
\end{figure*}

To show the importance of UE selection and power allocation, we first generate the results in the indoor setting for a simple centralized scheduler, i.e., round-robin scheduler with fixed maximum transmission powers in both directions. In the HD system (HD synchronous), in each direction, each cell selects UEs in a round-robin manner. In the FD system, in each timeslot, each cell chooses the same UE as selected in the HD system with a randomly selected UE for the other direction to make an FD pair. Figures~\ref{fig:rr_cdf}(a) and~\ref{fig:rr_cdf}(b) show the distribution of average downlink and uplink throughputs, for different BS self-interference cancellation capabilities. In the downlink direction, in most of the cases (70$\%$), there is no FD gain, which is due to the lack of any intelligent selection procedure during FD operation. In the uplink, due to the cancellation of self-interference, the FD system throughput is higher than the HD system. The difference improves with increased self-interference cancellation capability. From a system point of view, which includes both uplink and downlink, this round-robin scheduling does not provide sufficient FD capacity gain. This demonstrates the need for an intelligent scheduling algorithm to provide a gain during FD operation which can benefit both uplink and downlink. 

\begin{figure*} 
\centering
\includegraphics[width = 5.5in] {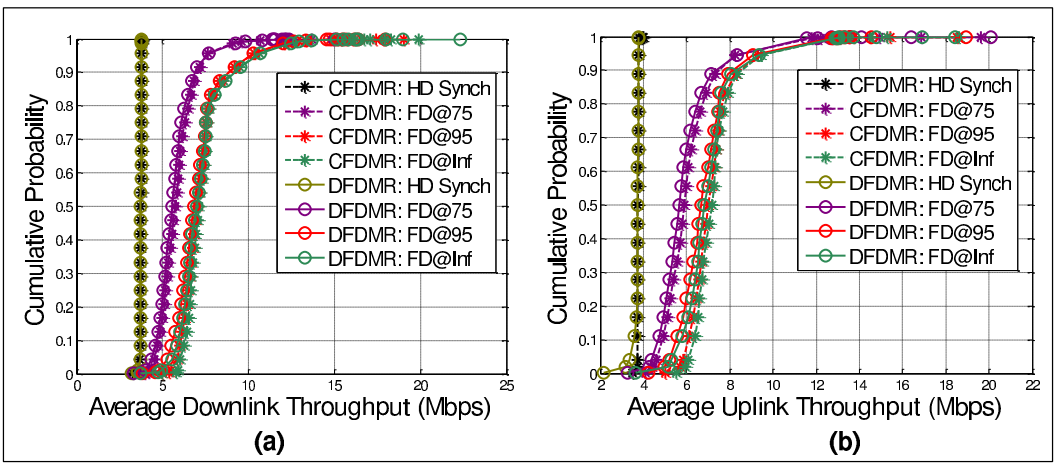}
\caption{Distribution of average data rates for the half-duplex system and full duplex system with both distributed and centralized scheduling algorithms in an indoor multi-cell scenario.}
\label{fig:cdf_indoor}
\end{figure*}

Next, we generate results with both the proposed distributed and centralized joint UE selection and power allocation procedure. Figures~\ref{fig:cdf_indoor}(a) and ~\ref{fig:cdf_indoor}(b) show the distribution of average downlink and uplink throughputs for both distributed and centralized methods. In this plot, the distribution is only shown for HD synchronous, FD@75, FD@95, and FD@Inf system to keep the plot readable, however Table~\ref{tab3} contains the average throughput over all UEs for all the simulated systems. It also contains the average throughput gain of FD systems compared to the HD synchronous system.

\begin {table*}
\caption {Average throughput (Mbps) over all UEs of half and full duplex systems with both distributed and centralized scheduling algorithms in an indoor multi-cell scenario. For a full duplex system, average throughput gain compared to the HD synchronous system is also given. } \label{tab3} 
\begin{center}
    \begin{tabular}{| l | l | l | l | l | l | l | l |}
    	\hline
		  &   \begin{tabular}[c]{@{}c@{}}\textbf{HD}\\ \textbf{Synchronous} \end{tabular} & \begin{tabular}[c]{@{}c@{}}\textbf{HD}\\ \textbf{Dynamic TDD} \end{tabular} & \textbf{FD@75} & \textbf{FD@85}  & \textbf{FD@95}  & \textbf{FD@105} & \textbf{FD@Inf} \\ \hline
		 \textbf {CFDMR: Downlink} & 3.75 & 3.77 & 5.85 $(56\%)$ & 6.76 $(80\%)$& 7.28 $(94\%)$& 7.39 $(97\%)$& 7.42  $(98\%)$\\ \hline
		 \textbf {DFDMR: Downlink} & 3.74 & 3.77 & 5.70 $(52\%)$& 6.49 $(74\%)$& 7.07 $(89\%)$& 7.25 $(94\%)$& 7.27 $(95\%)$\\ \hline
		 \textbf {CFDMR: Uplink} & 3.75 & 3.73 & 6.15 $(64\%)$& 6.85 $(83\%)$& 7.23 $(93\%)$& 7.35 $(96\%)$& 7.38 $(97\%)$\\ \hline
		 \textbf {DFDMR: Uplink} & 3.69 & 3.72 & 5.96 $(61\%)$& 6.53 $(77\%)$& 6.88 $(87\%)$& 7.02 $(90\%)$& 7.06 $(91\%)$ \\ \hline	
    \end{tabular}
\end{center}
\end{table*}

The HD system shows a narrow distribution centered near 4 Mbps in both downlink and uplink whereas the FD system shows a wider distribution since the scheduler takes advantage of the variable nature of the interference to assign FD operation with an appropriate data rate whenever possible. The dynamic TDD HD system has similar performance as the synchronous HD system since the same kind of channel model is assumed between different nodes, and therefor there is not much different in the interference experienced by a node in both systems. In this scenario, the distributed algorithm performs nearly as well as the centralized solution for almost all the systems. In general, the throughput gain of FD system compared the HD system increases as the self-interference cancellation improves. With the higher self-interference cancellation values, the FD system nearly doubles the capacity compared to the HD system. 
%Further, Table~\ref{tab3} shows the average improvement in the 5$\%$ cell edge capacity, which also increases as the self-interference cancellation increases.

From the simulation one can also observe the dependency between FD/HD operation selection in our scheduler and the self-interference cancellation capability, that is, the lower the self-interference cancellation, the fewer the number of cells in a timeslot that are scheduled in FD mode. This is verified by counting the average number of cells per timeslot which are in FD mode or HD mode or with no transmission as shown in Table~\ref{tab4}. With 75 dB self-interference cancellation, on average 84$\%$ of the cells operate in FD mode, while with 105 dB, 98$\%$ of the cells operate in FD mode. Note that in the HD system, in each timeslot, all cells transmit in one direction (either uplink or downlink). These results are for the centralized method; similar results are obtained for the distributed method.

\begin{table*}
\caption {Average number of cells per slot in different modes in an indoor multi-cell scenario.} \label{tab4} 
\begin{center}
\begin{tabular}{|c|c|c|c|c|c|c|}
\hline
\multicolumn{1}{|l|}{} & \begin{tabular}[c]{@{}c@{}}\textbf{HD}\\ (\textbf{Downlink}, \textbf{Uplink})\end{tabular} & \textbf{FD@75} & \textbf{FD@85} & \textbf{FD@95} & \textbf{FD@105} & \textbf{FD@Inf} \\ \hline
\textbf{FD Mode}                & -                                                      & 84\%  & 93\%  & 97\%  & 98\%   & 98\%   \\ \hline
\textbf{HD Mode}                & (100\%, 100\%)                               & 16\%  & 7\%   & 3\%   & 2\%    & 2\%    \\ \hline
\textbf{No Transmission}        & (0\%, 0\%)                                   & 0\%   & 0\%   & 0\%   & 0\%    & 0\%    \\ \hline
\end{tabular}
\end{center}
\end{table*}

To analyze the performance of FD operation in the outdoor scenario, as we mentioned earlier in Section~\ref{sec:DSaSP}, we first simulate the dense outdoor multi-cell scenario. In this case, the average throughput gain of the FD system is only 25$\%$ in the downlink and 32$\%$ in the uplink with the centralized scheduler. These gains do not vary with self-interference cancellation because strong inter-cell interference dominates the self-interference and decreases the opportunities for capacity improvement due to FD operation. These results show that it is not very beneficial to use FD radios in dense outdoor environments due to the high inter-cell interference. This observation motivates us to investigate the performance of FD radios in sparse outdoor environments.   

\begin{figure*}
\centering
\includegraphics[width = 5.5in] {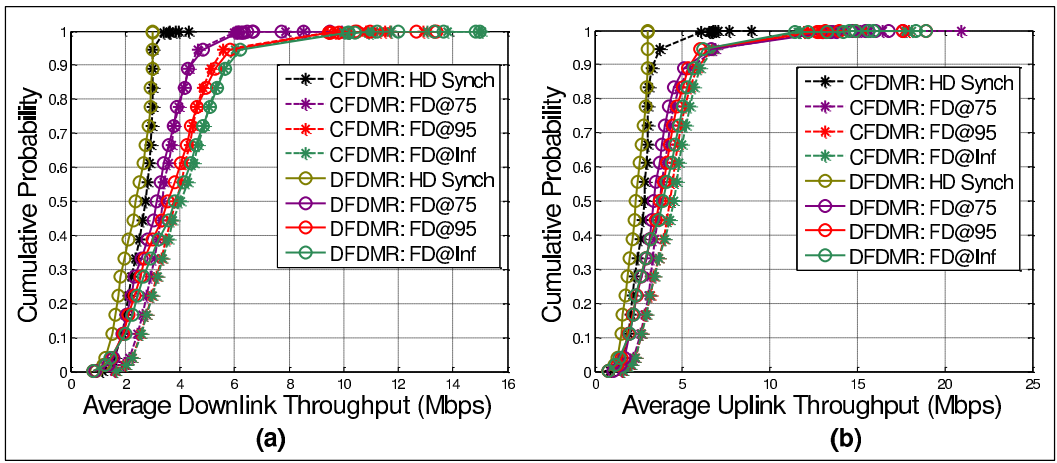}
\caption{Distribution of average data rates for the half-duplex system and full duplex system with both distributed and centralized scheduling algorithms in an outdoor multi-cell scenario.}
\label{fig:cdf_outdoor}
\end{figure*}

We simulate the sparse outdoor multi-cell scenario as shown in Figure~\ref{fig:fig5}(b). We run our simulation for several random drops of twelve picocells in a hexagonal cell with a width of 500 meters. Figures~\ref{fig:cdf_outdoor}(a) and~\ref{fig:cdf_outdoor}(b) show the distribution of average downlink and uplink throughputs, and Table~\ref{tab5} shows the average throughput over all UEs for all the systems and also the gain of the FD system as compared to the HD synchronous system. Similar to the indoor scenario, FD increases the capacity of the system significantly over the HD case, where the increase is proportional to the amount of self-interference cancellation. In this case also the distributed scheduling algorithm gives results close to the centralized algorithm. In this outdoor scenario, the average throughput of a UE is lower compared to the indoor case, but it is distributed over a wider range. Moreover, the throughput increase due to FD operation is less than what it was in the indoor case. The reason behind this is that the inter-cell interference between a BS and UEs in neighboring cells is much stronger that in the indoor scenario. 

\begin {table*}
\caption {Average throughput (Mbps) over all UEs of half and full duplex systems with both distributed and centralized scheduling algorithms in an outdoor multi-cell scenario. For a full duplex system, average throughput gain compared to the HD synchronous system is also given.} \label{tab5} 
\begin{center}
    \begin{tabular}{| l | l | l | l | l | l | l | l |}
    	\hline
		  &   \begin{tabular}[c]{@{}c@{}}\textbf{HD}\\ \textbf{Synchronous} \end{tabular} & \begin{tabular}[c]{@{}c@{}}\textbf{HD}\\ \textbf{Dynamic TDD} \end{tabular} & \textbf{FD@75} & \textbf{FD@85}  & \textbf{FD@95}  & \textbf{FD@105} & \textbf{FD@Inf} \\ \hline
		 \textbf {CFDMR: Downlink} & 2.52 & 2.78 & 3.36 $(34\%)$ & 3.57 $(42\%)$& 3.84 $(52\%)$& 4.04 $(60\%)$& 4.08  $(62\%)$\\ \hline
		 \textbf {DFDMR: Downlink} & 2.23 & 2.38 & 3.05 $(37\%)$& 3.29 $(47\%)$& 3.52 $(58\%)$& 3.73 $(67\%)$& 3.80 $(70\%)$\\ \hline
		 \textbf {CFDMR: Uplink} & 2.70 & 2.94 & 3.96 $(47\%)$& 4.16 $(54\%)$& 4.31 $(60\%)$& 4.39 $(63\%)$& 4.43 $(64\%)$\\ \hline
		 \textbf {DFDMR: Uplink} & 2.22 & 2.38 & 3.57 $(61\%)$& 3.63 $(64\%)$& 3.67 $(66\%)$& 3.77 $(70\%)$& 3.79 $(71\%)$ \\ \hline	
    \end{tabular}
\end{center}
\end{table*}

In this case, for the centralized algorithm, the uplink throughput is higher than the downlink throughput, which also increases the gap between the performance of the distributed and centralized performance in the uplink. In the centralized greedy UE selection algorithm, the utility to select a UE is the difference between the marginal utility of the UE and the loss in the marginal utility of the selected UEs in other cells due to new interference generated from the UE being considered. In case of downlink, for all the potential UEs in the cell being considered, the second term, i.e. interference generation (from their BS) to other cells will be constant, whereas, in the uplink, since the interference generation also depends on the location of the UE, both utility gain and utility decrement of other cells vary from UE to UE. This difference provides more degrees of freedom for the uplink UE selection and therefore manages uplink multi-cell interference better than downlink case. 

\begin{table*}
\caption {Average number of cells per slot in different modes in an outdoor multi-cell scenario.} \label{tab6}
\begin{center}
\begin{tabular}{|c|c|c|c|c|c|c|}
\hline
\multicolumn{1}{|l|}{} & \begin{tabular}[c]{@{}c@{}}\textbf{HD}\\ (\textbf{Downlink}, \textbf{Uplink})\end{tabular} & \textbf{FD@75} & \textbf{FD@85} & \textbf{FD@95} & \textbf{FD@105} & \textbf{FD@Inf} \\ \hline
\textbf{FD Mode}                & -                                                    & 36\%  & 50\%  & 56\%  & 57\%   & 57\%   \\ \hline
\textbf{HD Mode}                & (81\%, 88\%)                                                  & 62\%  & 48\%   & 42\%   & 41\%    & 41\%    \\ \hline
\textbf{No Transmission}        & (19\%, 12\%)                                                      & 2\%   & 2\%   & 2\%   & 2\%    & 2\%    \\ \hline
\end{tabular}
\end{center}
\end{table*}

Table~\ref{tab6} shows the average number of cells per slot which are in FD mode, HD mode or with no transmission with the centralized scheduling method. First of all, in the HD system, in contrast to the indoor scenario, we can see that some cells are not transmitting at all in some slots. This is due to the higher inter-cell interference between the BS and UEs in neighboring cells; the system throughput is higher when certain cells are not scheduled for transmission, resulting in reduced inter-cell interference. Further, for the same reason, the average number of cells operating in FD mode is smaller than the indoor scenario. In this case, the number of cells in FD mode also increases with self-interference cancellation.

\subsection{Full Duplex Gain for the Non-full Buffer Traffic Model}\label{sec:NFT}
In this section we analyze the performance of the FD system with non-full buffer FTP traffic~\cite{3GPP:3}. In this case, each UE has requests to download or/and upload files of 1.25 MB. The time interval between completion of a file transmission and an arrival of a new request is exponentially distributed with a mean of 1 second. The delay for each UE, which is defined as the total time it experiences from the request arrival to the completion of downloading or uploading a file is calculated. A significant delay improvement, due to simultaneous downloading and uploading in an FD system is observed as shown in Table~\ref{tab7}, which shows the average delay a UE experiences for different systems. Moreover, a UE downloads 48$\%$, 69$\%$, 83$\%$, 90$\%$, and 92$\%$ more files and uploads 56$\%$, 75$\%$, 86$\%$, 88$\%$, and 90$\%$ more files in the FD system compared to those in the HD system with 75 dB, 85 dB, 95 dB, 105 dB, and perfect self-interference cancellation, respectively.

\begin {table*} 
\caption {Average delay (Seconds) in an indoor multi-cell scenario.} \label{tab7} 
\begin{center}
    \begin{tabular}{| l | l | l | l | l | l | l |}
    	\hline
		 & \textbf{HD Synchronous} & \textbf{FD@75} & \textbf{FD@85}  & \textbf{FD@95}  & \textbf{FD@105} & \textbf{FD@Inf} \\ \hline
		\textbf {Downlink} & 2.43 & 1.33 & 1.05 & 0.89 & 0.83 & 0.81 \\ \hline
		\textbf {Uplink}  & 2.39 & 1.23 & 1.01 & 0.92 & 0.89 & 0.87 \\ \hline		
    \end{tabular}
\end{center}
\end{table*}

\subsection{Signaling Overhead}
In this section we compute the signaling overhead required to enable FD scheduling algorithms compared to the existing HD system. As mentioned in Section~\ref{sec:CEP}, in our FD system, each UE needs to send the channel measurement information of its neighborhood. In our simulations, we derive a threshold for each UE to determine inclusion in its potential strong interferer list for UE-to-UE interference. For an UE $u$, given its threshold, all other such UEs for which UE-to-UE channel is higher than the threshold will be considered as strong interferers, and UE $u$ will send the channel information for these UEs to its BS. A downlink UE gets interference from both neighboring BSs and uplink UEs. The channel measurement from the BSs is used to derive the threshold for the UE-to-UE channels for each UE. Each UE measures the channel with all its neighboring BSs and derives the average channel strength of its BS-to-UE interference channel. This average channel strength is used as the threshold for the UE-to-UE interference channel. Let us assume that on an average there are $K$ strong UE interferers. 

We assume the channel information is represented by 8-bits. If a UE sends this information every 2ms, which is the maximum periodic frequency of the SRS transmission of a UE \cite{DahlmanLTE}, the total overhead in each cell, would be $4KN_m$ kbps. In our simulations, in the indoor scenario, where $N_m =$ 8, and the average value of $K$ observed equals 7. The average overhead in the indoor scenario is thus 224 kbps. In the outdoor scenario, it is 320 kbps ($N_m =$ 10, $K = $ 8). For example, for a LTE system with 10 MHz bandwidth and 16 QAM, where the peak LTE uplink capacity is 25.5 Mbps \cite{3GPP:5},  the UE-to-UE channel measurement incurs less than 2$\%$ overhead. 

We also compare the signaling overhead of the distributed and centralized algorithms in terms of average outbound traffic generated by each BS. In the centralized method, the centralized scheduler needs to collect a large set of channel information from each BS, which includes, (1) channels with other BSs, (2) channels with all the UEs in the system, (3) strong UE-to-UE channels. It also needs to collect weights of all UEs. In this case, each BS generates $(M + M N_m + N_m K) \times 8$ bits per transmission time interval (TTI). In case of the distributed approach, each BS generates $(2 + 2 + 2M + K) \times 8$ bits during initialization and $(2+2) \times n_I \times 8$ bits during the iterative process, where $n_I$ is the number of iterations. In the case of the indoor system, based on our simulation results, if we assume $K = 7, n_I = 7$, then for the centralized approach each BS generates 1096 bits per TTI, and in the case of the distributed approach, each BS generates 456 bits per TTI.

\section{Conclusion}\label{sec:conc}
We investigated the application of common carrier FD radios to resource managed small-cell systems in a multi-cell deployment. Assuming FD capable BSs with HD UEs, we present a joint uplink and downlink scheduler which does UE selection and power allocation to maximize the network utility in a distributed manner. It operates in FD mode when conditions are favorable, and otherwise defaults to HD mode. The proposed distributed algorithm performs nearly as well as the centralized solution but with much lower signaling overhead. Our simulation results show that an FD system using a practical design parameter of 95 dB self-interference cancellation at each BS can improve the capacity by 90\% in an indoor multi-cell hot zone scenario and 60\% in an outdoor multi picocell scenario. From these results we conclude that in both indoor small-cell and sparse outdoor environment, FD base stations with an intelligent scheduling algorithm are able to improve capacity significantly with manageable signaling overhead.

\bibliographystyle{IEEEtran}
\bibliography{FD_references}

\begin{thebibliography}{10}
\providecommand{\url}[1]{#1}
\csname url@samestyle\endcsname
\providecommand{\newblock}{\relax}
\providecommand{\bibinfo}[2]{#2}
\providecommand{\BIBentrySTDinterwordspacing}{\spaceskip=0pt\relax}
\providecommand{\BIBentryALTinterwordstretchfactor}{4}
\providecommand{\BIBentryALTinterwordspacing}{\spaceskip=\fontdimen2\font plus
\BIBentryALTinterwordstretchfactor\fontdimen3\font minus
  \fontdimen4\font\relax}
\providecommand{\BIBforeignlanguage}[2]{{%
\expandafter\ifx\csname l@#1\endcsname\relax
\typeout{** WARNING: IEEEtran.bst: No hyphenation pattern has been}%
\typeout{** loaded for the language `#1'. Using the pattern for}%
\typeout{** the default language instead.}%
\else
\language=\csname l@#1\endcsname
\fi
#2}}
\providecommand{\BIBdecl}{\relax}
\BIBdecl

\bibitem{Cisco}
\BIBentryALTinterwordspacing
``{Cisco visual network index}: Forecast and methodology 2013-2018,'' Cisco
  white paper, June 2014. [Online]. Available: \url{www.cisco.com}
\BIBentrySTDinterwordspacing

\bibitem{Ericsson}
\BIBentryALTinterwordspacing
``Ericsson mobility report,'' June 2013. [Online]. Available:
  \url{www.ericsson.com}
\BIBentrySTDinterwordspacing

\bibitem{Horizon}
\BIBentryALTinterwordspacing
``Creating a smart network that is flexible, robust and cost effective,''
  Horizon 2020 Advanced 5{G} Network Infrastructure for Future Internet PPP,
  Industry Proposal (Draft Version 2.1), 2013. [Online]. Available:
  \url{http://www.networks-etp.eu/}
\BIBentrySTDinterwordspacing

\bibitem{Khandani10}
A.~K. Khandani, ``Methods for spatial multiplexing of wireless two-way
  channels,'' October 2010, {US} Patent 7,817,641.

\bibitem{Katti10}
J.~I. Choi, M.~Jain, K.~Srinivasan, P.~Levis, and S.~Katti, ``Achieving single
  channel, full duplex wireless communication,'' in \emph{Proceedings of the
  Sixteenth Annual International Conference on Mobile Computing and Networking
  (MOBICOM)}.\hskip 1em plus 0.5em minus 0.4em\relax ACM, 2010, pp. 1--12.

\bibitem{Knox12}
M.~Knox, ``Single antenna full duplex communications using a common carrier,''
  in \emph{Wireless and Microwave Technology Conference (WAMICON), 2012 IEEE
  13th Annual}, April 2012, pp. 1--6.

\bibitem{Katti13}
D.~Bharadia, E.~McMilin, and S.~Katti, ``Full duplex radios,'' in
  \emph{Proceedings of the ACM SIGCOMM 2013}.\hskip 1em plus 0.5em minus
  0.4em\relax ACM, 2013, pp. 375--386.

\bibitem{Duarte13}
M.~Duarte, A.~Sabharwal, V.~Aggarwal, R.~Jana, K.~Ramakrishnan, C.~Rice, and
  N.~Shankaranarayanan, ``Design and characterization of a full-duplex
  multiantenna system for {WiFi} networks,'' \emph{Vehicular Technology, IEEE
  Transactions on}, vol.~63, no.~3, pp. 1160--1177, March 2014.

\bibitem{survey_JSAC}
A.~Sabharwal, P.~Schniter, D.~Guo, D.~Bliss, S.~Rangarajan, and R.~Wichman,
  ``In-band full-duplex wireless: Challenges and opportunities,''
  \emph{Selected Areas in Communications, IEEE Journal on}, vol.~32, no.~9, pp.
  1637--1652, Sept 2014.

\bibitem{survey_kim}
D.~Kim, H.~Lee, and D.~Hong, ``A survey of in-band full-duplex transmission:
  From the perspective of {PHY} and {MAC} layers,'' \emph{Communications
  Surveys Tutorials, IEEE}, vol.~pp, no.~99, pp. 1--1, Feb 2015.

\bibitem{NGMN_5G}
\BIBentryALTinterwordspacing
``{NGMN} {5G} white paper,'' March 2015. [Online]. Available:
  \url{www.ngmn.org}
\BIBentrySTDinterwordspacing

\bibitem{kumu_5G}
S.~Hong, J.~Brand, J.~Choi, M.~Jain, J.~Mehlman, S.~Katti, and P.~Levis,
  ``Applications of self-interference cancellation in {5G} and beyond,''
  \emph{Communications Magazine, IEEE}, vol.~52, no.~2, pp. 114--121, Feb 2014.

\bibitem{SanjayCISS13}
S.~Goyal, P.~Liu, S.~Hua, and S.~Panwar, ``Analyzing a full-duplex cellular
  system,'' in \emph{Information Sciences and Systems (CISS), 2013 47th Annual
  Conference on}, March 2013, pp. 1--6.

\bibitem{SanjayICC14}
S.~Goyal, P.~Liu, S.~Panwar, R.~A. DiFazio, R.~Yang, J.~Li, and E.~Bala,
  ``Improving small cell capacity with common-carrier full duplex radios,'' in
  \emph{2014 IEEE International Conference on Communications (ICC)}, June 2014,
  pp. 4987--4993.

\bibitem{ChoiSTR12}
Y.-S. Choi and H.~Shirani-Mehr, ``Simultaneous transmission and reception:
  Algorithm, design and system level performance,'' \emph{Wireless
  Communications, IEEE Transactions on}, vol.~12, no.~12, pp. 5992--6010,
  December 2013.

\bibitem{XShen13}
X.~Shen, X.~Cheng, L.~Yang, M.~Ma, and B.~Jiao, ``On the design of the
  scheduling algorithm for the full duplexing wireless cellular network,'' in
  \emph{Global Communications Conference (GLOBECOM), 2013 IEEE}, Dec 2013, pp.
  4970--4975.

\bibitem{HyunICTC}
H.-H. Choi, ``On the design of user pairing algorithms in full duplexing
  wireless cellular networks,'' in \emph{Information and Communication
  Technology Convergence (ICTC), 2014 International Conference on}, Oct 2014,
  pp. 490--495.

\bibitem{multi_cell_GPonly}
T.~Nguyen, D.~Ngo, A.~Nasir, and J.~Khan, ``Utility-based interference
  management for full-duplex multicell networks,'' in \emph{Communications
  (ICC), 2015 IEEE International Conference on}, June 2015, pp. 1914--1919.

\bibitem{survey_WLAN}
K.~Thilina, H.~Tabassum, E.~Hossain, and D.~I. Kim, ``Medium access control
  design for full duplex wireless systems: challenges and approaches,''
  \emph{Communications Magazine, IEEE}, vol.~53, no.~5, pp. 112--120, May 2015.

\bibitem{survey_IC_downlink}
A.~Hamza, S.~Khalifa, H.~Hamza, and K.~Elsayed, ``A survey on inter-cell
  interference coordination techniques in {OFDMA}-based cellular networks,''
  \emph{Communications Surveys Tutorials, IEEE}, vol.~15, no.~4, pp.
  1642--1670, March 2013.

\bibitem{Barghi12}
S.~Barghi, A.~Khojastepour, K.~Sundaresan, and S.~Rangarajan, ``Characterizing
  the throughput gain of single cell {MIMO} wireless systems with full duplex
  radios,'' in \emph{Modeling and Optimization in Mobile, Ad Hoc and Wireless
  Networks (WiOpt), 2012 10th International Symposium on}, May 2012, pp.
  68--74.

\bibitem{DiINFOCOM}
B.~Di, S.~Bayat, L.~Song, and Y.~Li, ``Radio resource allocation for
  full-duplex {OFDMA} networks using matching theory,'' in \emph{Computer
  Communications Workshops (INFOCOM WKSHPS), 2014 IEEE Conference on}.\hskip
  1em plus 0.5em minus 0.4em\relax IEEE, 2014, pp. 197--198.

\bibitem{Shaocommletter}
S.~Shao, D.~Liu, K.~Deng, Z.~Pan, and Y.~Tang, ``Analysis of carrier
  utilization in full-duplex cellular networks by dividing the co-channel
  interference region,'' \emph{IEEE Communications Letters}, vol.~18, no.~6,
  pp. 1043--1046, June 2014.

\bibitem{GMSingleCell}
M.~Al-Imari, M.~Ghoraishi, P.~Xiao, and R.~Tafazolli, ``Game theory based radio
  resource allocation for full-duplex systems,'' in \emph{Vehicular Technology
  Conference (VTC Spring), 2015 IEEE 81st}, May 2015.

\bibitem{ouyang2015leveraging}
W.~Ouyang, J.~Bai, and A.~Sabharwal, ``Leveraging one-hop information in
  massive {MIMO} full-duplex wireless systems,'' \emph{arXiv preprint
  arXiv:1509.00539}, 2015.

\bibitem{Nguyenduplo}
D.~Nguyen, L.~Tran, P.~Pirinen, and M.~Latva{-}aho, ``On the spectral
  efficiency of full-duplex small cell wireless systems,'' \emph{CoRR}, vol.
  abs/1407.2628, 2014.

\bibitem{duplo_site}
\BIBentryALTinterwordspacing
{The Duplo Website}. [Online]. Available: \url{http://www.fp7-duplo.eu/}
\BIBentrySTDinterwordspacing

\bibitem{Simeone2014full}
O.~Simeone, E.~Erkip, and S.~Shamai, ``Full-duplex cloud radio access networks:
  An information-theoretic viewpoint,'' \emph{arXiv preprint arXiv:1405.2092},
  2014.

\bibitem{multi_cell_yun}
J.~Yun, ``Intra and inter-cell resource management in full-duplex heterogeneous
  cellular networks,'' \emph{Mobile Computing, IEEE Transactions on}, vol.~pp,
  no.~99, April 2015.

\bibitem{Goyal_CommMag}
S.~Goyal, P.~Liu, S.~Panwar, R.~A. DiFazio, R.~Yang, and E.~Bala, ``Full duplex
  cellular systems: {Will} doubling interference prevent doubling capacity?''
  \emph{Communications Magazine, IEEE}, vol.~53, no.~5, pp. 121--127, May 2015.

\bibitem{alves2014average}
H.~Alves, C.~H. de~Lima, P.~H. Nardelli, R.~Demo~Souza, and M.~Latva-aho, ``On
  the average spectral efficiency of interference-limited full-duplex
  networks,'' in \emph{9th International Conference on Cognitive Radio Oriented
  Wireless Networks and Communications (CROWNCOM)}.\hskip 1em plus 0.5em minus
  0.4em\relax IEEE, 2014, pp. 550--554.

\bibitem{Quekhybrid}
J.~Lee and T.~Q.~S. Quek, ``Hybrid full-/half-duplex system analysis in
  heterogeneous wireless networks,'' \emph{IEEE Transactions on Wireless
  Communications}, vol.~14, no.~5, pp. 2883--2895, May 2015.

\bibitem{Goyal_ICC16}
S.~Goyal, C.~Galiotto, N.~Marchetti, and S.~S. Panwar, ``Throughput and
  coverage for a mixed full and half duplex small cell network,'' in \emph{2016
  IEEE International Conference on Communications (ICC)}, May 2016.

\bibitem{DahlmanLTE}
E.~Dahlman, S.~Parkvall, and J.~Skold, \emph{{4G LTE / LTE-Advanced for Mobile
  Broadband}}.\hskip 1em plus 0.5em minus 0.4em\relax Oxford:Elsevier, 2011.

\bibitem{3GPP:2}
\BIBentryALTinterwordspacing
``Physical channels and modulation ({Release 10}),'' TS 36.211, v.10.5.0, June
  2012. [Online]. Available: \url{www.3gpp.org}
\BIBentrySTDinterwordspacing

\bibitem{3GPP:1}
\BIBentryALTinterwordspacing
``Further enhancements to lte time division duplex ({TDD}) for downlink-uplink
  ({DL-UL}) interference management and traffic adaptation,'' TR 36.828,
  v.11.0.0, June 2012. [Online]. Available: \url{www.3gpp.org}
\BIBentrySTDinterwordspacing

\bibitem{Stoylar05}
A.~L. Stolyar, ``On the asymptotic optimality of the gradient scheduling
  algorithm for multiuser throughput allocation,'' \emph{Operations Research},
  vol.~53, no.~1, pp. 12--25, Jan-Feb 2005.

\bibitem{PF_proof}
H.~Kim, K.~Kim, Y.~Han, and S.~Yun, ``A proportional fair scheduling for
  multicarrier transmission systems,'' in \emph{Vehicular Technology
  Conference, 2004. VTC2004-Fall. IEEE 60th}, vol.~1, Sept 2004, pp. 409--413.

\bibitem{jalali2000data}
A.~Jalali, R.~Padovani, and R.~Pankaj, ``Data throughput of {CDMA-HDR} a high
  efficiency-high data rate personal communication wireless system,'' in
  \emph{Vehicular Technology Conference Proceedings, IEEE 51st}, 2000, pp.
  1854--1858.

\bibitem{3GPP:5}
\BIBentryALTinterwordspacing
``Physical layer procedures ({Release 11}),'' TS 36.213, v11.0.0, September
  2012. [Online]. Available: \url{www.3gpp.org}
\BIBentrySTDinterwordspacing

\bibitem{D2DneighborDisSRS}
H.~Tang, Z.~Ding, and B.~Levy, ``Enabling {D2D} communications through neighbor
  discovery in {LTE} cellular networks,'' \emph{Signal Processing, IEEE
  Transactions on}, vol.~62, no.~19, pp. 5157--5170, Oct 2014.

\bibitem{D2DneighborDisDMRS}
K.~Lee, W.~Kang, and H.-J. Choi, ``A practical channel estimation and feedback
  method for device-to-device communication in {3GPP} {LTE} system,'' in
  \emph{Proceedings of the 8th International Conference on Ubiquitous
  Information Management and Communication (ICUIMC)}.\hskip 1em plus 0.5em
  minus 0.4em\relax ACM, 2014.

\bibitem{D2DneighborDisFlashLinq}
F.~Baccelli, N.~Khude, R.~Laroia, J.~Li, T.~Richardson, S.~Shakkottai,
  S.~Tavildar, and X.~Wu, ``On the design of device-to-device autonomous
  discovery,'' in \emph{Communication Systems and Networks (COMSNETS), 2012
  Fourth International Conference on}, Jan 2012, pp. 1--9.

\bibitem{3GPP:4}
\BIBentryALTinterwordspacing
``{X2} general aspects and principles ({Release 8}),'' TS 36.420, v.8.0.0,
  December 2007. [Online]. Available: \url{www.3gpp.org}
\BIBentrySTDinterwordspacing

\bibitem{boyd2007tutorial}
S.~Boyd, S.~J. Kim, L.~Vandenberghe, and A.~Hassibi, ``A tutorial on geometric
  programming,'' \emph{Optimization and engineering}, vol.~8, no.~1, pp.
  67--127, 2007.

\bibitem{chiang2007power}
M.~Chiang, C.~W. Tan, D.~Palomar, D.~O'Neill, and D.~Julian, ``Power control by
  geometric programming,'' \emph{Wireless Communications, IEEE Transactions
  on}, vol.~6, no.~7, pp. 2640--2651, July 2007.

\bibitem{venturino2009coordinated}
L.~Venturino, N.~Prasad, and X.~Wang, ``Coordinated scheduling and power
  allocation in downlink multicell {OFDMA} networks,'' \emph{Vehicular
  Technology, IEEE Transactions on}, vol.~58, no.~6, pp. 2835--2848, July 2009.

\bibitem{yu2010joint}
W.~Yu, T.~Kwon, and C.~Shin, ``Joint scheduling and dynamic power spectrum
  optimization for wireless multicell networks,'' in \emph{Information Sciences
  and Systems (CISS), 2010 44th Annual Conference on}, March 2010, pp. 1--6.

\bibitem{kiani2007maximizing}
S.~G. Kiani, G.~E. Oien, and D.~Gesbert, ``Maximizing multicell capacity using
  distributed power allocation and scheduling,'' in \emph{2007 IEEE Wireless
  Communications and Networking Conference}, March 2007, pp. 1690--1694.

\bibitem{koutsopoulos2006cross}
I.~Koutsopoulos and L.~Tassiulas, ``Cross-layer adaptive techniques for
  throughput enhancement in wireless {OFDM}-based networks,'' \emph{Networking,
  IEEE/ACM Transactions on}, vol.~14, no.~5, pp. 1056--1066, Oct 2006.

\bibitem{SanjayCentArxiv}
\BIBentryALTinterwordspacing
S.~Goyal, P.~Liu, S.~Panwar, R.~Yang, R.~A. DiFazio, and E.~Bala, ``Full duplex
  operation for small cells,'' \emph{CoRR}, vol. abs/1412.8708, 2014. [Online].
  Available: \url{http://arxiv.org/abs/1412.8708}
\BIBentrySTDinterwordspacing

\bibitem{3GPP:3}
\BIBentryALTinterwordspacing
``Further advancements for {E-UTRA} physical layer aspects ({Release 9}),'' TR
  36.814, v.9.0.0, March 2010. [Online]. Available: \url{www.3gpp.org}
\BIBentrySTDinterwordspacing

\end{thebibliography}

\begin{IEEEbiography}[{\includegraphics[width=1in,height=1.25in,clip,keepaspectratio]{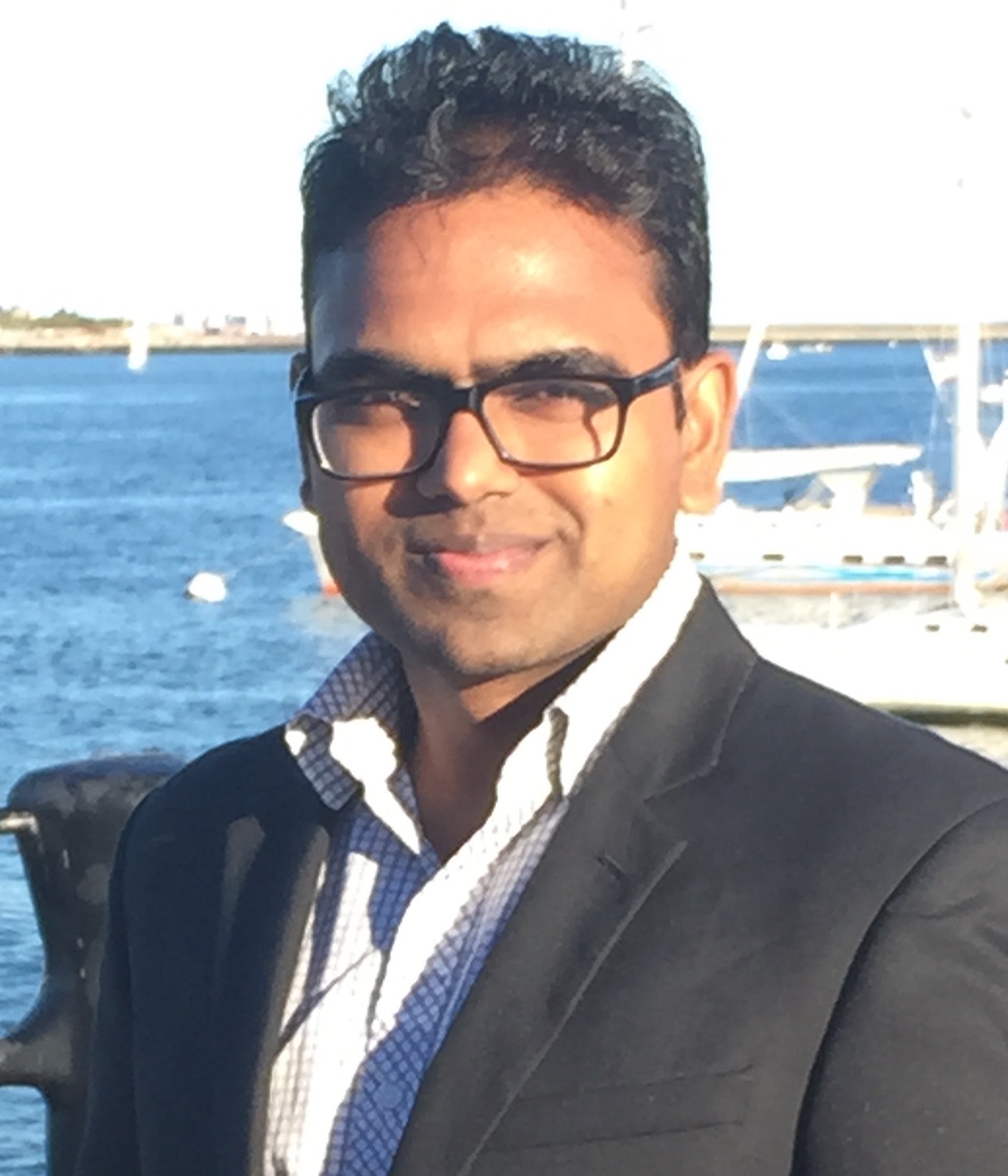}}]{Sanjay Goyal}
received his B.Tech. degree in communication and computer engineering from the LNM Institute of Information Technology, India, in 2009, and his M.S. degree in electrical engineering from
NYU Polytechnic School of Engineering, New York, in 2012. Currently, he is a Ph.D. candidate in the ECE Department at NYU Tandon School of Engineering. He was awarded, along with Carlo Galiotto, Nicola Marchetti, and Shivendra Panwar, the Best Paper Award in IEEE ICC 2016. His research interests are in designing and analyzing wireless network protocols with full duplex communication, especially for the MAC layer.
\end{IEEEbiography}  
\begin{IEEEbiography}[{\includegraphics[width=1.1in,height=1.25in,clip,keepaspectratio]{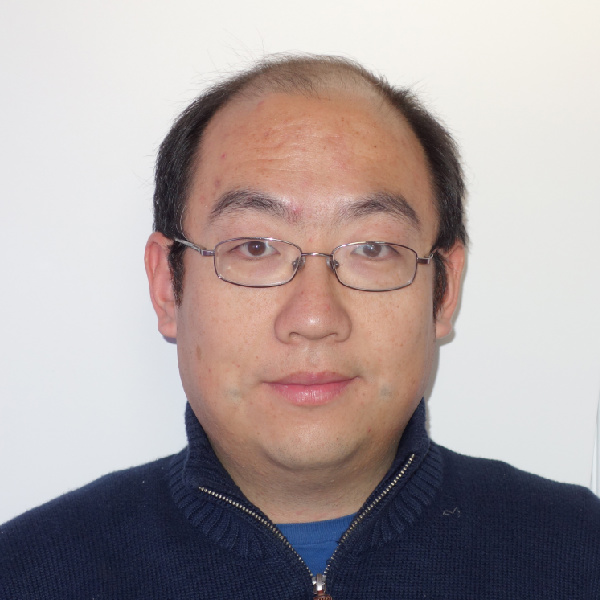}}]{Pei Liu}
is a research assistant professor of Electrical and Computer Engineering at NYU Tandon School of Engineering. He received his Ph.D. degree in Electrical and Computer Engineering from Polytechnic University in 2007. He received his B.S. and M.S. degrees in electrical engineering from Xi'an Jiaotong University, China, in 1997 and 2000, respectively. His research interests are in designing and analyzing wireless network protocols with an emphasis on cross-layer optimization, especially with the PHY and MAC layers. Currently, his research topics include wireless communications, wireless networks, and video over wireless.
\end{IEEEbiography} 
\begin{IEEEbiography}[{\includegraphics[width=1in,height=1.25in,clip,keepaspectratio]{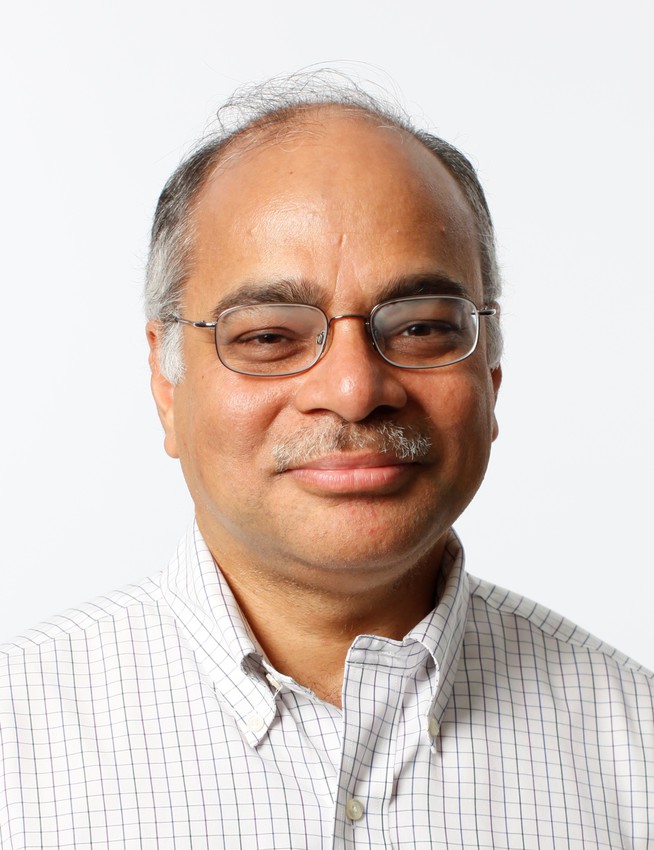}}]{Shivendra S. Panwar}
[F] is a Professor in the Electrical and Computer Engineering Department at NYU Tandon School of Engineering. He received the B.Tech. degree in electrical engineering from the Indian Institute of Technology Kanpur, in 1981, and the M.S. and Ph.D. degrees in electrical and computer engineering from the University of Massachusetts, Amherst, in 1983 and 1986, respectively. He is currently the Director of the New York State Center for Advanced Technology in Telecommunications (CATT), the Faculty Director of the NY City Media Lab, and member of NYU WIRELESS. He spent the summer of 1987 as a Visiting Scientist at the IBM T.J. Watson Research Center, Yorktown Heights, NY, and has been a Consultant to AT\&T Bell Laboratories, Holmdel, NJ. His research interests include the performance analysis and design of networks. Current work includes wireless networks, switch performance and multimedia transport over networks. He is an IEEE Fellow and has served as the Secretary of the Technical Affairs Council of the IEEE Communications Society. He is a co-editor of two books, Network Management and Control, Vol. II, and Multimedia Communications and Video Coding, both published by Plenum. He has also co-authored TCP/IP Essentials: A Lab based Approach, published by the Cambridge University Press. He was awarded, along with Shiwen Mao, Shunan Lin and Yao Wang, the IEEE Communication Society's Leonard G. Abraham Prize in the Field of Communication Systems for 2004. He was also awarded, along with Zhengye Liu, Yanming Shen, Keith Ross and Yao Wang, the Best Paper in 2011 Multimedia Communications Award.
\end{IEEEbiography}

\end{document}